%% file: OAM.tex
\documentclass[lettersize,journal]{IEEEtran}
\IEEEoverridecommandlockouts
\usepackage{cite}
\usepackage{amsmath,amssymb,amsfonts}
\usepackage{algorithmic}
\usepackage{graphicx}
\usepackage{textcomp}
\usepackage{xcolor}
\def\BibTeX{{\rm B\kern-.05em{\sc i\kern-.025em b}\kern-.08em
    T\kern-.1667em\lower.7ex\hbox{E}\kern-.125emX}}

\usepackage{array}
\usepackage[caption=false,font=normalsize,labelfont=sf,textfont=sf]{subfig}
\usepackage{stfloats}
\usepackage{url}
\usepackage{verbatim}
\usepackage{graphicx}
\usepackage{cite}
\usepackage{siunitx}
\usepackage{algorithm}
\usepackage{caption}
\usepackage{subcaption}
\usepackage{subfiles}
\usepackage{color}
\usepackage{url}
\usepackage{verbatim}
\usepackage{commath}
\usepackage{stackengine,mathtools}
\usepackage{blindtext}
\usepackage [acronym] {glossaries}
\usepackage{tikz}
\usepackage {pbox}
\usepackage{balance}
\usepackage[none]{hyphenat}
\usepackage{xurl}
\usepackage{xr}
\usepackage[normalem]{ulem}
\usepackage{newtxtext}
\usepackage[shortlabels]{enumitem}
\usepackage{titlesec}

\DeclareMathOperator*{\argmin}{arg\,min}

\newcommand{\name} {\mbox{OrthoVortex}}

\begin{document}

\title{Fast Vortex Beam Alignment for OAM Mode Multiplexing in LOS MIMO Networks}

\author{
Poorya Mollahosseini,~\IEEEmembership{Graduate Student Member,~IEEE,} and, Yasaman Ghasempour,~\IEEEmembership{Member,~IEEE.}
\IEEEcompsocitemizethanks
{
\IEEEcompsocthanksitem This work has been submitted to the IEEE for possible publication. Copyright may be transferred without notice, after which this version may no longer be accessible.
\IEEEcompsocthanksitem This work was funded by the National Science Foundation (grant CNS-2145240), the Qualcomm Innovation Fellowship, and the Air Force Office of Scientific Research (award FA9550-24-1-0144).
\IEEEcompsocthanksitem Authors are with the Department of Electrical and Computer Engineering, Princeton University, Princeton, NJ 08540 USA. E-mail: poorya@princeton.edu, ghasempour@princeton.edu.
}
}

\maketitle

\begin{abstract}
\input{Sections/Abstract}

\end{abstract}


\vspace{-0.2cm}
\section{Introduction}
\input{Sections/Intro}

\vspace{-0.2cm}
\section{Related Work}
\input{Sections/RelatedWork}

\vspace{-0.2cm}
\section{Primer}
\input{Sections/Primer}

\vspace{-0.2cm}
\section{\name's Design}
\input{Sections/Design}

\vspace{-0.2cm}
\section{Implementation and Experimental Setup}
\input{Sections/Platform}

\vspace{-0.2cm}
\section{Evaluation}
\input{Sections/Evaluation}

\vspace{-0.25cm}
\section{Discussion and Limitations}
\input{Sections/Discussions}

\vspace{-0.3cm}
\section{Conclusion}
\input{Sections/Conclusion}

\vspace{-0.3cm}
\bibliographystyle{IEEEtran}
\balance
\bibliography{References.bib,OAM.bib}

\end{document}

%% file: Sections/Abstract.tex
Orbital Angular Momentum (OAM)-based communication systems offer high-capacity multiplexing in line-of-sight (LOS) scenarios; yet, their performance is sensitive to nodal misalignment, which disrupts modal orthogonality, hindering the data multiplexing gain. To tackle this challenge, we present \name, a novel framework that estimates the misalignment angles and applies the appropriate phase correction to restore orthogonality between modes. Unlike purely theoretical prior efforts that rely on impractical fully digital arrays or exhaustive beam scans, \name~introduces and leverages the \textit{cross-modal phase}, as a unique signature for identifying the misalignment angles. \name~is a few-shot alignment technique, making it feasible for real-world implementations. Our key contributions include: \textit{(i)} a robust angle estimation and phase correction framework based on the physics of OAM propagation that estimates the misalignment and restores modal orthogonality, \textit{(ii)} the first-ever experimental validation of OAM beam alignment with RF transceivers, and \textit{(iii)} a comprehensive analysis of practical constraints, including the impact of antenna count and bandwidth. Simulations and over-the-air measurements using low-cost, rapidly prototyped metasurfaces operating at 120 GHz demonstrate that \name~achieves fast and precise misalignment estimation (mean absolute error of \ang{0.69} for azimuth and \ang{2.54} for elevation angle). Further, \name~can mitigate the inter-modal interference, yielding more than 12 dB increase in signal-to-interference ratio and more than 4.5-fold improvement in link capacity.

%% file: Sections/Intro.tex
\begin{figure}[t]
    \centering
    \includegraphics[width=0.36\textwidth]{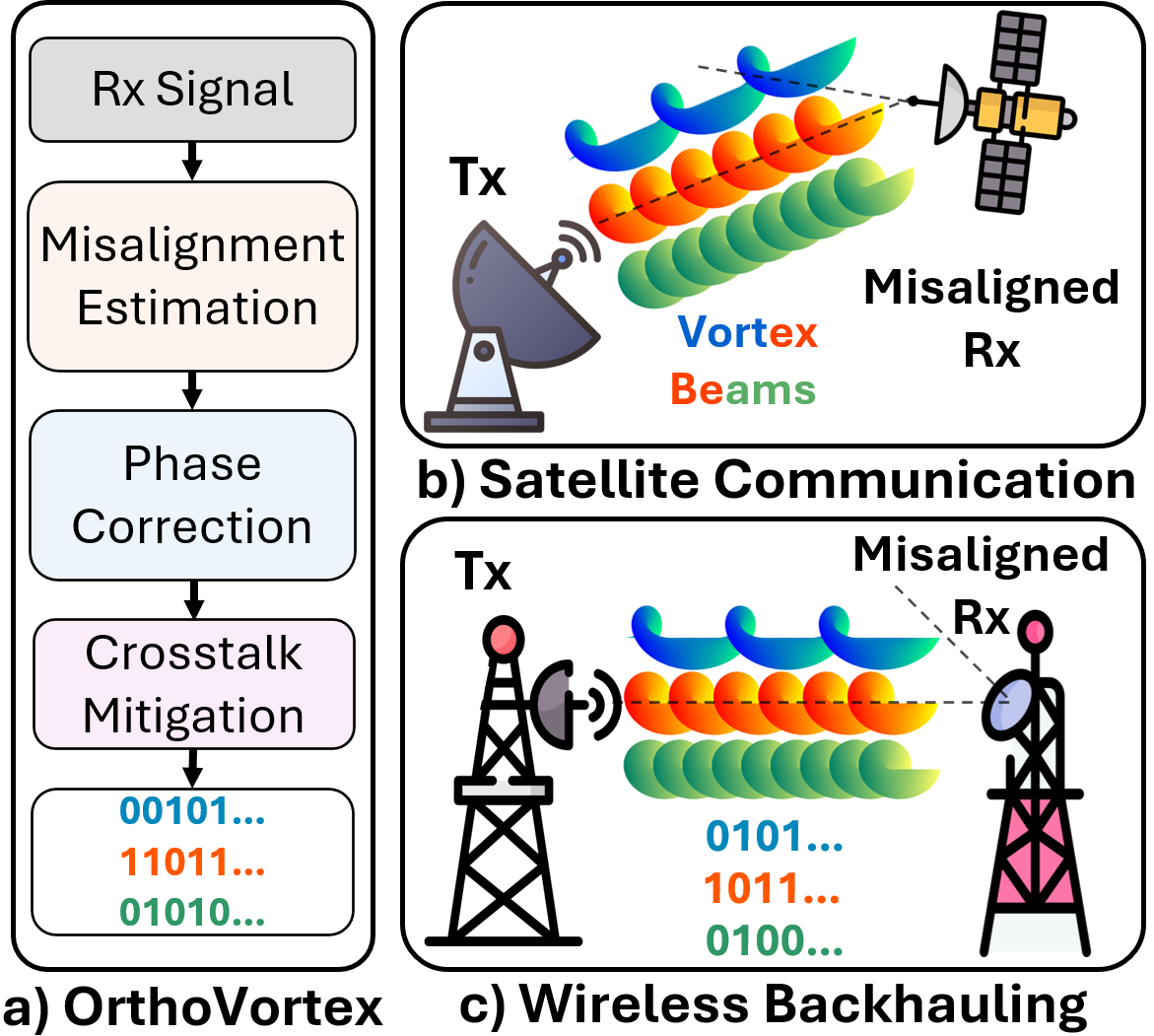}
    \vspace{-0.1cm}
    \caption{\textnormal{ (a) \name~estimates the misalignment angles and uses them to restore modal orthogonality, enhancing mode multiplexing in OAM links. This approach has potential applications in (b) satellite-to-ground communications and (c) wireless backhauling.}} 
    \vspace{-0.5cm}
    \label{fig:intro}
\end{figure}

In recent years, the pursuit of higher data rates has shifted focus to underutilized millimeter-wave (mmWave) and sub-terahertz (sub-THz) bands, where wider bandwidths are available. In parallel, multiple-input-multiple-output (MIMO) systems have also been recognized as a key technology to scale data rates through spatial multiplexing. However, the spatial multiplexing gain in such systems depends heavily on multipath-rich environments~\cite{telatar1999capacity}. 

Unfortunately, higher-frequency channels are often characterized by their line-of-sight (LOS) dominance and sparsity (due to the higher diffraction loss), hindering the performance of conventional spatial multiplexing. Further, new wireless applications have emerged, including wireless backhauling, aerial communication, and wireless satellite communication, in which the wireless channel is inherently sparse. 

To achieve spatial multiplexing gain in these settings, the literature has explored the use of beams that carry orbital angular moments (OAM)~\cite{mahmouli20134,ren2017line}. From an electromagnetic (EM) perspective, OAM refers to a property whereby the wavefront exhibits a helical or spiral structure, forming what is known as a \textit{vortex beam}. This beam carries angular momentum about its propagation axis, with the OAM mode number $l$ indicating the number of phase twists around the axis~\cite{willner2021orbital}.

Crucially, OAM beams with different modes are inherently orthogonal, preventing mutual interference even when transmitted simultaneously over the same frequency. Therefore, different OAM modes can carry independent data streams at the same time, also called \textit{mode multiplexing}, and boost the achievable data rate~\cite{trichili2019communicating}. In other words, this inherent mode orthogonality enables spatial multiplexing in LOS channels, where traditional single-user MIMO (SU-MIMO) techniques falter due to the absence of multipath diversity. There have been impressive demonstrations of OAM's potential for data multiplexing, such as those in~\cite{yagi2021field, lee2018experimental, sasaki2023demonstration, hirabe201940}, where links achieving 137 Gbps over 100~m and 117 Gbps over 200~m were established.

However, despite the fundamental physics of OAM being understood for decades and the aforementioned experimental demonstrations of it, we have yet to see its widespread adoption. One of the main barriers to realizing OAM communication systems in practice is the sensitivity to alignment. Misalignment between transceivers disrupts the unique helical phase structure of OAM beams. As a result, the received modes lose their inherent orthogonality, inducing inter-modal interference (IMI), which manifests itself as crosstalk between data streams. This crosstalk severely degrades the multiplexing gain in OAM communications.

In this paper, to make these vortex beams orthogonal again, we present \name, the first framework that achieves fast OAM beam alignment and enables mode multiplexing in LOS SU-MIMO communications. In practice, the misalignment between the transmitter (Tx) and receiver (Rx) arrays introduces spatially varying phase variations across the receiver aperture, which distorts the OAM helical phase profile, yielding significant inter-modal interference. Such distortions can be, in principle, canceled by applying appropriate counteracting phase shifts. This correction effectively \textit{flattens} the distorted phase front, realigning it with the ideal helical structure and restoring modal orthogonality. However, calculating this phase correction mask depends on accurate estimation of the misalignment angles (elevation and azimuth), a non-trivial task in practice.

Conventional solutions fail to address this challenge. Algorithms such as MUSIC require fully digital arrays to sample the impinging phase gradients, an impractical requirement at mmWave and sub-THz frequencies due to prohibitive hardware costs and power consumption~\cite{yang2018digital}. Exhaustive beam scanning, another traditional approach, sweeps through a predefined grid of angles to identify optimal alignment. However, this method introduces latency proportional to the grid resolution, making it incompatible with mobile applications where real-time adaptation is critical~\cite{zhu2014demystifying}. These limitations underscore a critical gap: the absence of a hardware-efficient, low-latency method to estimate misalignment angles and enable phase correction in dynamic environments. 

To address this gap, \name~exploits the fundamental properties of OAM beams to estimate the misalignment angles accurately and in a few-shot manner. Fig.~\ref{fig:intro}a depicts the overview of \name. Our key insight is that the relative received phase across different OAM beams, which we introduce as the \textit{cross-modal phase}, captures critical information about misalignment angles—enabled by the helical nature of the OAM phase profile. By exploiting this property, \name~achieves few-shot angle estimation without requiring fully digital arrays or exhaustive scanning. The estimated angles are then used to compute a phase correction mask, restoring orthogonality and re-enabling multiplexing gains.

\noindent
\textbf{Contributions:}
First, we derive a closed-form mathematical model that describes the received signal at a misaligned receiver, grounded in the physics of OAM wave propagation. This model serves as the foundation for our angle estimation framework. Second, we discover that the received signal can be decomposed into two distinct components: a \textit{mode-dependent term} and a \textit{mode-independent term}. This decomposition allows us to estimate the misalignment angles by observing the signal across multiple modes, without requiring additional information such as the distance between the transmitter and receiver. Third, we demonstrate that frequency diversity can be exploited to enhance system performance at zero additional time overhead. Finally, we validate \name~through extensive simulations and over-the-air experiments. We generate OAM beams using low-cost, rapidly prototyped metasurfaces operating at 120 GHz. Our results demonstrate a precise angle estimation, achieving a mean absolute error (MAE) of \ang{0.69} for the azimuth and \ang{2.54} for the elevation angle. Further, \name~offers a significant enhancement in canceling the IMI in OAM communication links, improving the signal-to-interference ratio (SIR) at the receiver by 12 dB and increasing the channel capacity by a factor of 4.5. We emphasize that this is the \textit{first-ever experimental work} that addresses the challenge of misalignment in radio frequency (RF) OAM communication links.

{\name~is particularly useful in scenarios where the communication nodes are intended to remain aligned—such as known-trajectory systems like ground-to-satellite links (Fig.~\ref{fig:intro}b) or static deployments for wireless backhauling (Fig.~\ref{fig:intro}c)—but may still experience unintentional misorientation due to environmental factors like gravitational effects or wind. Additionally, by enabling real-time angle correction, \name~serves as a critical step toward extending OAM multiplexing to mobile applications, which have so far remained unexplored due to alignment sensitivity in OAM-based systems}. By bridging the gap between theoretical OAM advantages and practical implementation, \name~paves the way to make high-capacity LOS OAM networks a reality for 6G and beyond. 

The remainder of the paper is structured as follows. Section~\ref{sec:related} discusses prior works. Section~\ref{sec:primer} provides an overview of OAM fundamentals. Section~\ref{sec:design} details the modeling behind \name's phase correction and angle estimation framework. Section~\ref{sec:setup} outlines the experimental setup, and Section~\ref{sec:eval} presents the evaluation results. Section~\ref{sec:discussion} discusses limitations and potential directions for future work. Finally, Section~\ref{sec:conclusion} concludes the paper.

%% file: Sections/RelatedWork.tex
\label{sec:related}
\noindent
\textbf{Beam Misalignment in the Optical Domain.}
Since OAM multiplexing was first studied in the optical regime, the problem of misalignment has also been explored in this domain. Proposed solutions include beacon-less tracking systems~\cite{li2018experimental}, methods exploiting the gradient of the intensity profile~\cite{xie2015exploiting, xie2017localization}, and SVD-based approaches~\cite{pang2020experimental}. These solutions assume the incident electric profile is given; hence, several image processing techniques can be used to extract misalignment information and compensate for it. Unfortunately, capturing the incident electric field at sufficient resolution for this approach to work would require massive, fully digital arrays that are power-demanding and costly (if at all possible) in the mmWave/sub-THz bands.

\noindent
\textbf{Beam Misalignment in the RF Domain.}
OAM multiplexing in misaligned cases has also attracted attention in the RF domain. The impact of misalignment on channel capacity has been studied in~\cite{cui2024effect}, while efforts to restore modal orthogonality include beam steering~\cite{chen2018beam, zhengjuan2021broadband, chen2019reception, yu2022uca}, and IMI cancellation~\cite {saito2021efficient}. However, these works assume the misalignment angles are known and focus solely on IMI mitigation, limiting their practical applicability. The second line of research focuses on estimating misalignment parameters~\cite{cheng2019achieving, jian2021non, chen2023index, vahidiniai2021array, tian2016beam, sun2023enhanced, chen2020multi, long2021aoa}. These approaches, however, require complete knowledge of the MIMO channel matrix and have only been validated through simulations, lacking experimental verification. 
To the best of our knowledge, the closest work to \name~with experimental analysis is~\cite{gao2019misalignment}, where a vector network analyzer was used to measure the S-parameter at 20 GHz and infer the misalignment parameters based on processing the distribution of the phase. However, the main limitation is the need for phase extraction over a large 2-D plane with sub-wavelength sampling resolution, which would demand a fully digital massive MIMO array. 

Our work distinguishes itself from prior efforts in three key ways: \textit{(i)} we are the first to experimentally demonstrate misalignment angle estimation through end-to-end sub-THz over-the-air transmission and reception, \textit{(ii)} our approach does not require a fully digital array for angle estimation or IMI mitigation, and \textit{(iii)} finally, \name~estimates misalignment angles in a few-shot manner, making it a practical solution for vortex beam alignment in practice compared to exhaustive or even hierarchical beam scanning protocols that incur much larger overheads.

%% file: Sections/Primer.tex
\label{sec:primer}

\subsection{Fundamentals of OAM}
\label{sec:primer1}
It was discovered in 1992 that EM waves can carry OAM~\cite{allen1992orbital} in addition to spin angular momentum, which is also referred to as polarization. While polarization is a property of the wave's electric field direction, OAM is a property of the wave's phase structure. Specifically, OAM beams feature a helical phase front described by $e^{il\phi}$, where $i$ is the imaginary unit, $\phi$ is the azimuthal angle, and $l$ is the OAM mode, an integer describing the number of $2\pi$ phase shifts in the phase profile of the beam. This helical phase front results in a phase singularity at the beam's center, which yields a donut-shaped intensity profile. The intensity distribution of OAM beams is linked to the mode number $l$. Higher-order OAM modes (larger $l$) exhibit greater divergence, meaning the beam spreads more rapidly as it propagates~\cite{willner2021orbital}. 

\begin{figure}[t]
    \centering
    \includegraphics[width=0.47\textwidth]{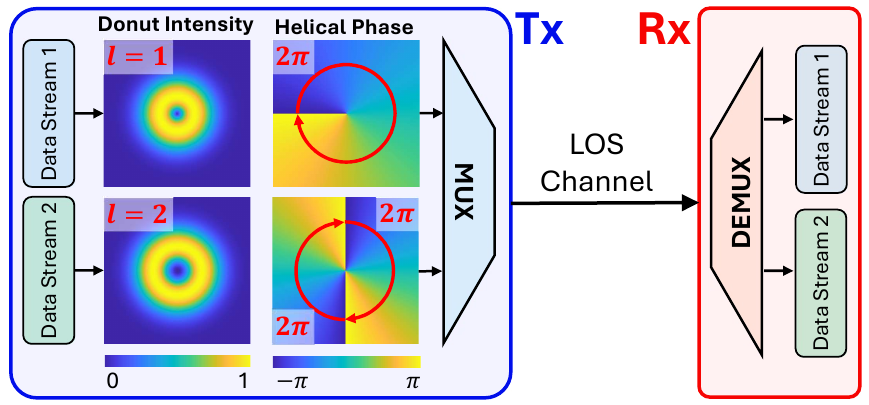}
    \vspace{-0.1cm}
    \caption{{OAM beams are characterized by a donut-shaped intensity profile with a central null and a helical phase front that rotates $l$ times around the singularity point. Different OAM modes can be used simultaneously to carry independent data streams without interference.}} 
    \label{fig:OAM}
    \vspace{-0.5cm}
\end{figure}

The key property of OAM beams is their inherent modal orthogonality: OAM modes with different integer mode numbers $l$ are orthogonal to each other~\cite{willner2021orbital}. In other words, the inner product of two OAM field profiles with different modes ($l$) integrates to zero over a $2\pi$ azimuthal period. This property enables the simultaneous transmission of multiple OAM beams at different modes without interference, making OAM a powerful tool for multiplexing independent data streams {between a base station and a single user with multiple antennas}. This property, along with the phase and intensity profiles of several OAM modes, is shown in Fig.~\ref{fig:OAM}.

Similar to conventional beam steering, OAM beams can also be generated with an array of antennas connected to phase shifters~\cite{mohammadi2009orbital}. For instance, a uniform circular array (UCA) of antennas with a progressive phase shift of $l\phi$ can generate an OAM mode $l$~\cite{gong2017generation}. Similarly, 2D planar arrays can also be exploited to generate OAM beams~\cite{bhardwaj2019generation}. Hence, OAM multiplexing can be realized with existing multi-antenna hardware. Other techniques for OAM generation in the literature involve the use of low-cost spiral phase plates~\cite{wei2015generation} or metasurfaces~\cite{qin2018transmission} that add a helical phase profile to the incident wave.

\subsection{OAM as a Spatial Multiplexing Technique in LOS Channels}
\label{sec:primer2}
From a theoretical standpoint, OAM mode multiplexing is a special case of {spatial multiplexing in single-user MIMO (SU-MIMO) scenarios}, and thus shares the same capacity upper bound established in classical MIMO theory~\cite{edfors2011orbital}. At mmWave and sub-THz frequencies, where the channel becomes increasingly LOS dominant due to reduced scattering and multipath, the total degrees of freedom are primarily governed by the geometry of the Tx and Rx arrays and the link distance. {This propagation behavior makes OAM particularly well-suited to such environments, as its orthogonality properties are highly sensitive to multipath and instead perform best under LOS or free-space conditions.}

Under these favorable conditions, OAM mode multiplexing can approach the theoretical capacity limit. Specifically, when both the Tx and Rx are perfectly aligned UCAs in a pure LOS channel, the optimal precoding and decoding matrices obtained from the singular value decomposition (SVD) of the channel matrix align with distinct OAM modes~\cite{edfors2011orbital}. In this case, OAM mode multiplexing becomes equivalent to performing spatial multiplexing with an SVD-optimal solution.

Even though OAM does not increase the upper bound of MIMO multiplexing gain, it offers a practical and low-complexity solution to achieve the said upper bound in LOS channels. The reason is two-fold: First, OAM-based transmission in a well-aligned LOS scenario does not require estimation of channel state information (CSI), i.e., OAM mode multiplexing can be considered a \textit{CSI-free} spatial multiplexing technique. In contrast, in LOS-dominant channels, the classical SVD-based MIMO pre-coding becomes very sensitive to small amounts of error in channel sounding, as such errors get pronounced in matrix inversion, hindering the multiplexing gain. Second, unlike conventional MIMO, which requires complex digital signal processing for spatial stream separation (e.g., eigenmode beamforming), OAM beams can be separated with relatively simple mode-matching antennas~\cite{gil2021comparison}. However, these benefits come with a drawback: \textit{the sensitivity of OAM transmission to Tx-Rx alignment.}

\subsection{Challenge of Misalignment in OAM Links}
\label{sec:primer3}
While the science of OAM beams has been around for decades, we have yet to see its widespread adoption. The reason is rooted in the fact that while OAM modes are theoretically orthogonal under ideal conditions, practical implementations often face challenges due to misalignment between the Tx and Rx.

In a perfectly aligned system, the transmitter and receiver are ideally positioned such that \textit{(i)} the Rx array is perfectly parallel to the Tx array and \textit{(ii)} they share the same central axis, meaning the center of the transmitted OAM beam coincides exactly with the center of the receiver aperture. When these conditions are satisfied, the helical phase structure of the OAM mode is preserved, and the receiver can perfectly separate the modes by applying a conjugate phase pattern onto the received signal. 

Unfortunately, even slight misalignment can significantly degrade the orthogonality of OAM modes. The breakdown of modal orthogonality results in severe IMI, which manifests as crosstalk between the transmitted data streams. This interference degrades the SIR and the channel capacity, ultimately reducing the overall system performance. So the key question is: \textit{how to restore the modal orthogonality in misaligned cases in real-time?}

%% file: Sections/Design.tex
\label{sec:design}

\textit{Overview.} To address the challenge of misalignment in OAM-based communication systems, which disrupts modal orthogonality and diminishes multiplexing gains, we present \name. The insight behind the beam alignment strategy in \name~is that the phase distortion caused by misalignment is a pure function of the geometric features of the transmitter and receiver, and hence, it can be compensated through additional phase manipulation. In doing so, modal orthogonality will be restored, and thereby we can retain the multiplexing gain of OAM even in misaligned scenarios. 

However, such a phase correction requires accurate and real-time knowledge of the receiver's orientation relative to the transmitter. Conventionally, fully digital antenna arrays could be leveraged for precise angle of arrival estimation (which indicates the Rx's orientation) by using well-established techniques such as MUSIC. 
However, they are not scalable due to their high cost, energy consumption, and complexity, especially at higher frequencies. 

Instead, \name~assumes a fully connected hybrid MIMO architecture with multiple RF chains and many antenna elements, which is considered the most practical implementation of large arrays, striking a balance between complexity and performance. Further, exhaustive beam scanning could be used to identify misalignment angles, yet such methods are too slow for mobile scenarios where fast adaptation is critical. 

Instead, \name~leverages the unique helical phase front in the OAM beams to estimate the misalignment parameters with a few-shot measurements. Specifically, our key insight is that the \textit{cross-modal phase}, i.e., the relative phase of the received signal across several modes, serves as a fingerprint that can uniquely describe the Rx's orientation (relative to the Tx). This information is then used to compute a phase correction mask that cancels the distortion caused by the misalignment and restores modal orthogonality. This approach mitigates IMI and re-enables the multiplexing gain of OAM, ensuring high-capacity communication even in dynamic, mobile environments. Next, we describe the key components of \name.

\subsection{Restoring Modal Orthogonality via Phase Correction}
\label{sec:restore}

\begin{figure}[t]
    \centering
    \includegraphics[width=0.47\textwidth]{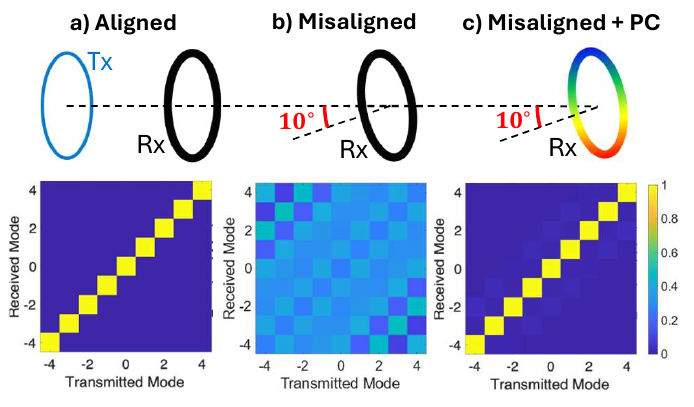}
    \vspace{-0.1cm}
    \caption{(a) When the Tx-Rx pair is aligned with each other, the modal orthogonality is preserved. (b) A slight misalignment leads to inter-modal interference. (c) The modal orthogonality is restored through appropriate phase correction (PC).} 
    \label{fig:misalignment}
    \vspace{-0.5cm}
\end{figure}

To investigate how modal orthogonality can be restored, we must first understand why misalignment causes its breakdown. {Generally, misalignment can be broken down into two components: \textit{translational} and \textit{angular}. Translational misalignment occurs when the Tx beam is not centered on the Rx aperture, meaning the center of the transmitted OAM beam does not align with the center of the receiver. Angular misalignment, on the other hand, occurs when the Tx and Rx arrays are not parallel due to a tilt in the Rx array.}

{This paper focuses on scenarios where the Tx is properly aligned, and only the Rx is misaligned. Such situations arise when the Tx either knows the Rx's trajectory \textit{a priori}, or when the Rx remains static but may be misoriented. In these cases, translational misalignment can be corrected by the Tx through beam steering—either mechanically or electronically. For instance, in satellite-to-ground communications, a ground station tracks a satellite’s known path and steers the beam accordingly. However, the satellite may still rotate due to unpredictable forces such as gravity gradients, atmospheric drag, or magnetic torques~\cite{starin2011attitude}. Similarly, in wireless backhaul links, wind-induced motion can rotate antenna arrays on towers. Therefore, unless otherwise stated, we use ``misalignment'' to refer specifically to Rx angular misalignment throughout this paper.}

Now that we have established the definition of misalignment in the scope of our work, let us dive deeper into understanding why misalignment causes the breakdown of OAM mode orthogonality. At a high level, when the Tx-Rx pair is perfectly aligned, all signals from the Tx to the Rx antennas travel the same distance. This uniformity ensures that the phase shifts due to signal propagation remain consistent across the entire Rx aperture, preserving the helical phase structure. However, misalignment introduces path length differences between the Tx and Rx antennas: when the Rx is tilted, signals reaching one side of its aperture traverse a longer path than those on the other side. These path differences result in spatially varying phase shifts across the receiver, distorting the helical phase profile such that it no longer matches the ideal $e^{il\phi}$ pattern. Consequently, the orthogonality between the modes breaks down. To restore orthogonality, we must cancel out these spatially varying phase shifts caused by misalignment.

Fortunately, if the orientation of the receiver with respect to the transmitter is known, these path-length differences can be mathematically modeled and compensated for through phase adjustments at the receiver. In the far field, the absolute distance between the Tx and Rx becomes irrelevant, as it introduces a uniform phase shift across the Rx aperture. Only the relative orientation, specifically, the elevation and azimuth angles, determines the phase structure that must be corrected. More specifically, if the center of the Tx is located at angles $(\theta, \phi)$ from the perspective of the Rx, the Rx can calculate the additional phase shift that it has to apply to its element located at $(x,y)$ as 
\begin{equation}
    \label{eq:phaseRx}
    P_{\text{rx}}(x,y) = -k \sin \theta (x \cos \phi + y \sin \phi),
\end{equation}
where $P_{\text{rx}}(x,y)$ is the additional phase. To demonstrate how misalignment leads to IMI and how it can be mitigated through phase correction, we conducted a simulation. Initially, we configured the Tx-Rx pair to be perfectly aligned, after which the Rx was tilted by \ang{10} in the elevation plane. For the tilted configuration, we tested two scenarios: with and without the proposed phase correction. The results are visualized in Fig.~\ref{fig:misalignment}. In this figure, we use a matrix representation to show the power spread between transmitted and received modes. Each cell in the matrix is color-coded, with brighter colors indicating higher power. In an ideally aligned system, power from a transmitted mode is fully retained in the corresponding received mode, resulting in a bright diagonal line and dark off-diagonal cells. This indicates the absence of leakage to other modes or zero IMI. Misalignment disrupts this orthogonality, scattering power into off-diagonal cells, which manifests as IMI. From Fig.~\ref{fig:misalignment}a, we observe that modal orthogonality holds when the Tx-Rx pair is perfectly aligned. However, even a slight misalignment, as shown in Fig.~\ref{fig:misalignment}b, causes power from one mode to leak into others, resulting in crosstalk. Fortunately, the knowledge of the tilt angle is sufficient to apply appropriate phase adjustments (according to Eq.~\eqref{eq:phaseRx}) and restore modal orthogonality, as demonstrated in Fig.~\ref{fig:misalignment}c. 

\textit{In summary, we need to estimate the misalignment angles to mitigate the IMI and benefit from the multiplexing gain of the OAM beams.} Next, we illustrate our system model and how \name~provides a framework for acquiring the necessary information about the angular orientation by leveraging the unique helical phase front of the OAM beams. 

\subsection{System Model}
\label{sec:geo}

\begin{figure}[t]
    \centering
    \includegraphics[width=0.47\textwidth]{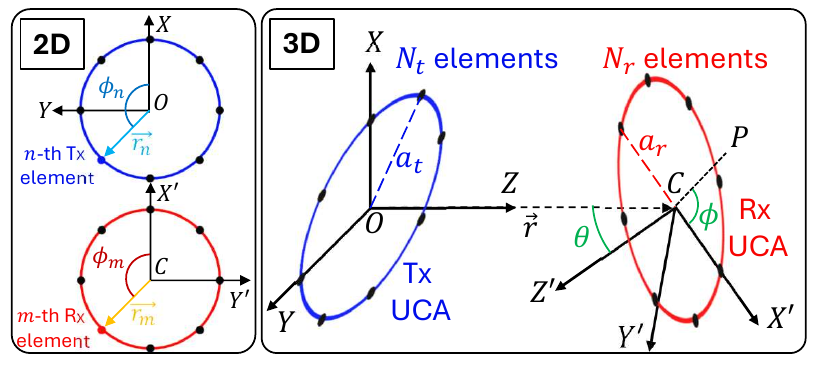}
    \vspace{-0.1cm}
    \caption{System Model. The Tx and Rx UCAs with their respective coordinate systems and parameters.} 
    \label{fig:overview}
    \vspace{-0.5cm}
\end{figure}

As discussed in Section~\ref{sec:restore}, estimating the misalignment angles is critical for applying the necessary phase corrections at the Rx to restore modal orthogonality and mitigate IMI. Therefore, we begin by introducing the specifications for the Tx and Rx and their respective parameters.

\noindent
\textbf{Antenna Architecture.} We assume that Tx and Rx are each equipped with a UCA of antennas, as their circular symmetry inherently aligns with the cylindrical phase structure of the OAM modes. We emphasize that OAM beams can also be generated with 2D planar arrays, and our mathematical modeling for angle estimation in \name~can be extended to such cases. As shown in Fig.~\ref{fig:overview}, the Tx and Rx UCAs are defined in separate coordinate systems $XYZ$ and $X'Y'Z'$, respectively. The Tx UCA comprises $N_t$ elements uniformly distributed with radius $a_t$ in the $XY$-plane, centered at the origin $O$. The position of the $n$-th Tx element is specified by its azimuthal angle $\phi_{n} = \frac{2\pi n}{N_t}$ and position vector $\vec{r}_n = a_t ( \cos\phi_n \hat{x} + \sin\phi_n \hat{y})$, where $0 \leq n \leq N_t-1$. Similarly, the Rx UCA contains $N_r$ elements with radius $a_r$ in the $X'Y'$-plane, centered at point $C$. The $m$-th Rx element, located at azimuthal angle $\phi_{m} = \frac{2\pi m}{N_r}$, has position vector $\vec{r}_m = a_r ( \cos\phi_m \hat{x'} + \sin\phi_m \hat{y'})$, where $0 \leq m \leq N_r-1$.

\noindent
\textbf{Transceiver Positions in the 3D Space.} Let $\vec{r} = r \hat{r}$ denote the vector from the center of the Tx array ($O$) to the center of the Rx array ($C$). In the Tx coordinate system, point $C$ is located at Cartesian coordinates $(0, 0, r)$. From the perspective of the Rx coordinate system, the Tx center $O$ appears at spherical coordinates $(r, \theta, \phi)$. In this frame, the vector pointing from $C$ to $O$ is simply $-\vec{r}$. Here, $\theta$ represents the angle between $-\vec{r}$ and the $Z'$-axis, while $\phi$ denotes the angle between the projection of $-\vec{r}$ onto the $X'Y'$-plane (labeled as $CP$ in Fig.~\ref{fig:overview}) and the $X'$-axis. Accordingly, in the Rx coordinate system, $-\vec{r}$ can be written as $-\vec{r} = r(\cos \phi \sin \theta\, \hat{x'} + \sin \phi \sin \theta\, \hat{y'} + \cos \theta\, \hat{z'})$.

\begin{figure}[t]
    \centering
    \vspace{-0.1cm}
    \includegraphics[width=0.47\textwidth]{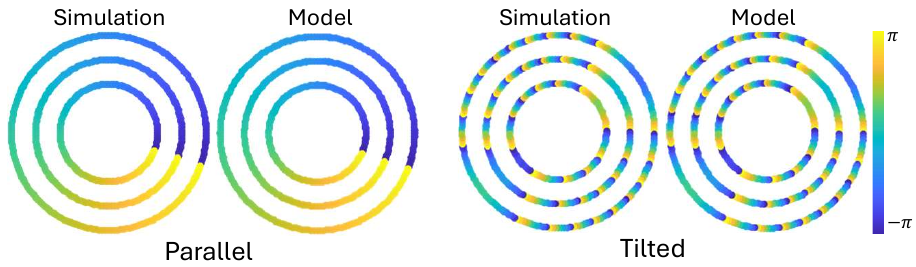}
    \caption{The phase distribution computed from full-wave EM simulation (left column in each pair) is compared with the output of the analytical model (right column).}
    \label{fig:EqMatch}
    \vspace{-0.5cm}
\end{figure}

Since our ultimate goal is to estimate the misalignment angles $\theta$ and $\phi$ based on the received signal at the Rx, we first derive a mathematical model that relates these angles to the signals captured at each Rx antenna, as a function of the transmitted waveform.

\noindent
\textbf{Signal Model.} Assume the Tx is transmitting a signal corresponding to the OAM mode $l$ at wavenumber $k$. This means that the $n$-th element of the Tx transmits the signal $e^{il\phi_{n}}$. In this case, assuming an isotropic single element, the signal component at the $m$-th receiver antenna, for mode $l$ and wavenumber $k$, denoted by $s_{m,l,k}$, can be formulated as:
\begin{equation}
    \label{Eq:E1}
    s_{m,l,k} = \frac{\alpha}{k} \sum_{n=0}^{N_t - 1} {e^{il\phi_n} 
     \frac{e^{-ik|\vec{r} - \vec{r}_n + \vec{r}_m|}}{|\vec{r} - \vec{r}_n + \vec{r}_m|}},
\end{equation}
where $\alpha $ is a scalar value that models all the constants, including the radiation pattern of a single element, total transmit power, etc. Eq. \eqref{Eq:E1} can be further simplified in several steps:
\begin{align} 
    \label{eq2}
    & s_{m,l,k} \approx \frac{\alpha}{k} \frac{1}{r}
    \sum_{n=0}^{N_t-1} e^{il\phi_n} 
    e^{-ik|\vec{r} - \vec{r}_n + \vec{r}_m|}
    \\ 
    \label{eq3}
    &\approx \frac{\alpha}{k} \frac{1}{r} 
    \sum_{n=0}^{N_t-1} 
    e^{il\phi_n} 
    e^{-ikr} 
    e^{-ik \hat{r} \cdot \vec{r}_m} 
    e^{ik \hat{v}_m \cdot \vec{r}_n}
    \\
    &= \frac{\alpha}{k} \frac{e^{-ikr}}{r}  e^{ika_r\sin(\theta)\cos(\phi-\phi_m)}
    \sum_{n=0}^{N_t-1} e^{il\phi_n} e^{ik \hat{v}_m \cdot \vec{r}_n}
    \\
    \label{eq5}
    &\approx 
    \frac{\alpha}{k} 
    \underbrace{\frac{e^{-ikr}}{r} 
    e^{i k a_r \sin(\theta) \cos(\phi - \phi_{m})}}_{\text{mode-independent}}
     N_t
    \underbrace{e^{i l (\delta_m + \gamma)} 
    J_l\left(\frac{k a_r a_t \rho_m}{r}\right)}_{\text{mode-dependent}},
\end{align}
where $ \vec{v}_m = \vec{r} + \vec{r}_m $, and $ \gamma $ is a constant term that depends solely on the orientation of the Rx.\footnote{$\gamma$ can be calculated as $\gamma = \tan^{-1}(w_2/w_1)$, where $\vec{w} = [w_1,w_2,w_3]^T = \hat{z'} \times \hat{z}$. Note that $\hat{z'}$ depends on the Rx's orientation and is unknown before $\theta$ and $\phi$ are estimated. Therefore, $\gamma$ is also unknown to the Rx and is a function of $\theta$ and $\phi$.} Here, $ J_l $ is the Bessel function of the first kind of order $ l $, $ \rho_m = \sqrt{\cos^2(\theta)\cos^2(\phi - \phi_m) + \sin^2(\phi - \phi_m)} $, and
\begin{equation}
    \label{eq:deltam}
    \delta_{m} = \tan^{-1}\left(\frac{\tan(\phi - \phi_m)}{\cos(\theta)}\right).
\end{equation}

Note that we used the amplitude approximation for the far-field to derive Eq.~\eqref{eq2}, and the phase approximation for the far-field to derive Eq.~\eqref{eq3}. {More specifically, the far-field assumption implies that the link distance is much greater than the dimensions of the transmit and receive arrays, i.e., $r \gg a_t$ and $r \gg a_r$. Under this condition, the amplitude variation across the array becomes negligible, allowing the approximation $|\vec{r} - \vec{r}_n + \vec{r}_m| \approx r$. For the phase, the assumption allows a first-order Taylor expansion of the propagation distance, leading to the approximation $|\vec{v}_m - \vec{r}_n| \approx v_m - \hat{v}_m . \vec{r}_n$.} Furthermore, the derivation of Eq.~\eqref{eq5} relies on the number of transmit antennas $N_t$ being sufficiently large such that the summation over them can be approximated by an integral, allowing it to be expressed in terms of a Bessel function.

This derivation shows that the received signal at each Rx antenna depends on the specific antenna chosen ($m$), the transmitted mode ($l$) and wavenumber ($k$), the geometric features of the Tx/Rx pair such as the radii of their UCAs ($a_r, a_t$), which are known, and the unknown parameters, including the misalignment angles ($\theta, \phi$) and the Tx-Rx distance ($r$).

We validated our mathematical model by comparing its output against a full-wave EM simulator. As an illustrative example, Fig.~\ref{fig:EqMatch} shows the phase of the received signal at the Rx for a setup with the following parameters: transmitter radius $a_t = 3$ cm, number of Tx elements $N_t = 160$, OAM mode $l = 1$, frequency $f = 120$ GHz, and link distance $r = 100$ m. The Rx consists of three concentric rings with radii 2, 3, and 4 cm, each populated with 120, 160, and 200 elements, respectively. Two scenarios are considered: one where the Rx is perfectly aligned with the Tx (i.e., $\theta = 0^\circ$, “Parallel” case), and one where the Rx is misaligned (tilted) with angles $\theta = 17.9^\circ$ and $\phi = -34.2^\circ$. The left two plots in Fig.~\ref{fig:EqMatch} correspond to the aligned case, while the right two plots represent the misaligned case. In each scenario, we compare the phase pattern obtained from EM simulation (left) with the one predicted by our analytical model (right). The phase results are plotted in a circular format to directly reflect the physical geometry of the Rx array, which consists of concentric circular rings. A close match is observed in both alignment conditions, verifying that our model accurately captures the phase distribution of the received signal across the Rx aperture. This confirms that Eq.~\eqref{eq5} is a valid and reliable representation of the received signal’s phase structure. Notably, this equation serves as the foundation for the angle estimation algorithm presented later in \name.

\subsection{Few-Shot Angle Estimation Framework}
\label{sec:framework}

\begin{figure*}[t]
    \centering
    \includegraphics[width=0.95\textwidth]{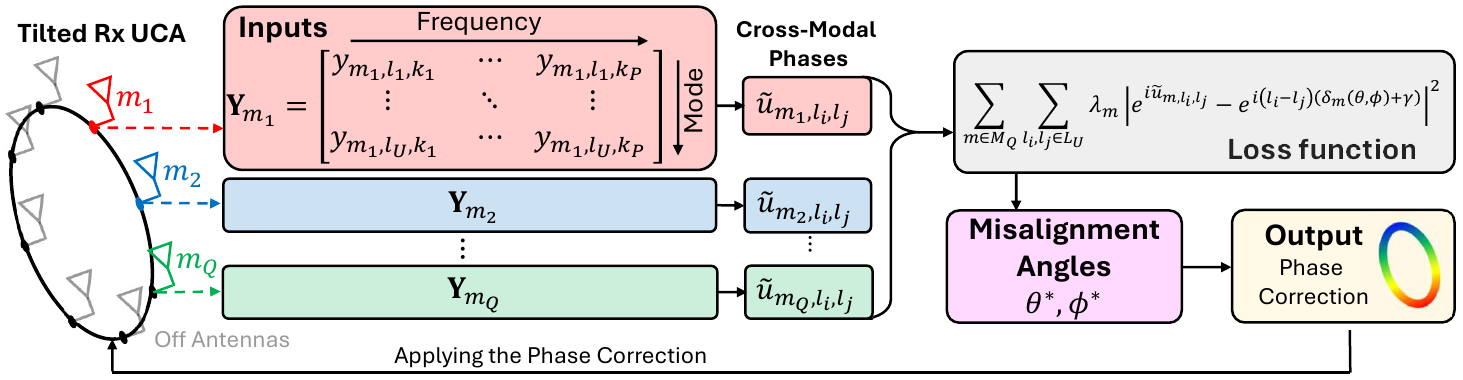}
    \caption{Signal Processing Pipeline. Each antenna independently estimates the cross-modal phase by observing the signal over different subcarriers and modes. These cross-modal phases are used to formulate the loss function according to Eq.~\eqref{eq:lossfunc}. By minimizing the loss, we estimate the misalignment angles and use them to calculate the phase correction mask, which is applied at the Rx to restore mode orthogonality.}
    \label{fig:angleEst}
\end{figure*}

In this section, we explain how to jointly estimate the misalignment angles $\theta$ and $\phi$ through a few-shot measurement of the received signal. Our key insight is that the relative phase of the received signal over distinct OAM modes, i.e., the \textit{cross-modal phase}, contains information about the misalignment angles. Specifically, our signal modeling in Eq.~\eqref{eq5} indicates that the unknown angles $(\theta,\phi)$ play a key role in the phase of the received signal, i.e., $s_{m,l,k}$. This impact can be broken down into two components: a \textit{mode-dependent} term and a \textit{mode-independent} term. The mode-independent term is a complex function of wavenumber, antenna geometry, Tx-Rx distance, and, of course, the unknown misalignment angles $(\theta,\phi)$. Accurate (sub-wavelength level) distance between Tx-Rx may not be known, hindering the estimation of $(\theta,\phi)$ through phase measurements. Interestingly, this term remains constant when different OAM modes are transmitted. Hence, \name~can cancel this term out by observing the received signal under different OAM modes. 

First, we highlight that when the argument and order of a Bessel function are real, the Bessel function is real-valued. Hence, the contribution of the Bessel term to the overall measured phase can either be $0$ or $\pi$. Second, the term $\delta_m$ is a direct function of $(\theta, \phi)$ as also indicated by Eq.~\eqref{eq:deltam} and thereby can be utilized for angle extraction.\footnote{To make this dependency more clear, we will use the notation $\delta_m(\theta,\phi)$ interchangeably with $\delta_m$.} To this end, \name~harnesses the unique properties of OAM wavefront and isolates $\delta_m$ by looking at the measured phase when distinct OAM modes are transmitted. In particular, we define $u_{m,l_2,l_1,k}$ as the phase difference between the received signals of OAM modes $l_2$ and $l_1$ at the $m$-th antenna element, which we write as
\begin{align}
    \nonumber
    u_{m,l_2,l_1,k} 
    & = \frac{1}{2} \angle\left(s_{m,l_2,k} \times s^{*}_{m,l_1,k}\right)^2 
    \\
    \label{eq:u1}
    & =  (l_2-l_1) (\delta_m(\theta,\phi)+\gamma),
\end{align}
where $\angle$ denotes the phase, and $^{*}$ denotes conjugate. The value of $u_{m,l_2,l_1,k}$ can be measured by isolating the $m$-th element and sampling the signal at two time slots, during which the transmitter sends two distinct OAM modes $l_1$ and $l_2$. We note that $l_1$ and $l_2$ are known \textit{a priori}. Also, note that the phase contribution of the Bessel function is canceled out when squaring the term $s_{m,l_2,k} \times s^{*}_{m,l_1,k}$.

From Eq.~\eqref{eq:u1}, we observe three unknowns, i.e., $\theta,\phi$, and $\gamma$. Therefore, \name~captures $u_{m,l_2,l_1,k}$ for three values of $m$ (corresponding to three different receiver antennas) to form and solve a system of three equations to determine $\theta$, $\phi$, and $\gamma$, the latter of which is not important to us.\footnote{These equations are independent if $|\phi_{m_i} - \phi_{m_j}| \neq \pi$ for all $i \neq j$, ensuring no two selected antennas are diametrically opposed on the Rx UCA.}  

{Although \name's angle estimation framework is based on phase information, which is inherently wrapped modulo $2\pi$, this phase wrapping does not lead to multiple optimal points in the search space. This is because we restrict the search range for the angles}: specifically, we enforce $0 \leq \theta < \pi/2$, ensuring that the front of the Rx array is facing toward the Tx. Within this range, $\theta$ is uniquely determined, eliminating ambiguity in its estimation. However, $\phi$ retains a potential $\pi$-ambiguity. To address this, the Rx computes two candidate phase masks, one using $\phi$ and the other using $\phi + \pi$, and applies both. It then selects the estimate that results in the higher received power after correction. 

\textit{Hence, six measurements—across two OAM modes and three receive elements—are required to jointly estimate the two misalignment angles.} \name~operates with just a single RF chain; however, when multiple RF chains are available—as is often the case in {fully connected hybrid MIMO} architectures designed to support spatial multiplexing—these phase measurements can be acquired in parallel. This parallelization leads to a linear reduction in the time complexity of \name.

So far, we have not considered the effect of noise in our framework. However, in real-world implementations, noise is an integral part of the system and will introduce errors into the angle estimation process. The main challenge is that noise corrupts our estimation of the cross-modal phase, which manifests itself as errors in the estimation of misalignment angles. In the presence of noise, we can model the received signal $y_{m,l,k}$ as
\begin{equation}
    \label{eq:ysn}
    y_{m,l,k} = s_{m,l,k} + n_{m,l,k},
\end{equation}
where $ n_{m,l,k} $ represents the additive white Gaussian noise component. To address this challenge, we employ two key strategies:

\textit{(i)} Leveraging multiple subcarriers.  

\textit{(ii)} Adding redundancy by utilizing the spatial diversity in Rx antennas or by using more OAM modes.

First, we observe a key property of the cross-modal phase: it is independent of frequency, as also evident from Eq.~\eqref{eq:u1}. This implies that the phase difference can be measured at any frequency, with all measurements yielding the same result. Therefore, multiple subcarriers can be used to improve resilience toward noise without imposing additional time overhead for the angle estimation framework in \name. Hence, extending Eq.~\eqref{eq:u1} for multiple subcarriers we can write:
\begin{align}
    \tilde{u}_{m, l_i, l_j} & = \frac{1}{2} \angle \left[ \sum_{k \in K_P} (y_{m,l_i,k} \times y^{*}_{m,l_j,k})^2 \right]
    \nonumber
    \\ &
    \label{eq:tildeu_m}
    \approx  
    (l_i-l_j)(\delta_m(\theta,\phi) + \gamma), 
\end{align}
where $K_P$ is a subset of size $P$ of the set of all available subcarriers, and $\tilde{u}_{m, l_i, l_j}$ denotes the average cross-modal phase across the subcarriers for modes $l_i$ and $l_j$ seen at the $m$-th receive antenna.

Second, transmitting more than two OAM modes and extracting the phase from more than three antennas can both add redundancy, which helps with noise mitigation. In particular, such redundancies would form an over-constrained system of equations (i.e., more equations than unknowns). The optimal solution minimizes the average residual error between the measured and modeled cross-modal phase. This approach inherently reduces the influence of individual estimation errors, improving the robustness of the angle estimation process.

Putting all pieces together, we can form an optimization framework that finds the desired angles $(\theta, \phi)$ by minimizing the distance between measured and modeled cross-modal phases over a set of OAM modes and received antennas, as follows:
\begin{multline}
    \label{eq:lossfunc}
    (\theta^*,\phi^*,\gamma^*) = \argmin_{(\theta,\phi,\gamma)} \\
    \sum_{m \in M_Q}  
    \sum_{l_i,l_j \in L_U}
    {
    \lambda_{m} 
    |
    e^{i\tilde{u}_{m, l_i, l_j}}
    -
    e^{i(l_i-l_j)(\delta_{m}(\theta,\phi) + \gamma)} 
    |^2
    },
\end{multline}
where $L_U$ is a subset of size $U$ of the set of all OAM modes and $M_Q$ is a subset of size $Q$ of the set of all Rx antennas. In other words, $U$ and $Q$ are the number of modes and Rx antennas used in the angle estimation framework. We note that increasing $U$ and $Q$ would improve the accuracy, but it will also inevitably increase the estimation overhead. 

Finally, not all antennas provide equally reliable estimates of the cross-modal phase, as the donut-shaped intensity profile of OAM beams might deliver higher received power on certain antennas and lower on others. To account for this, we introduce a weighting term $\lambda_{m}$, where $\lambda_{m}$ is a positive, non-decreasing function of the signal amplitude at the $m$-th antenna. This hyperparameter can be adjusted to optimize performance.

The signal processing pipeline of \name~is illustrated in Fig.~\ref{fig:angleEst}. In summary, the signals received at total of $Q$ antennas are sampled across $U$ OAM modes and $P$ frequencies, forming matrices $\mathbf{Y}_{m_1}, \dots, \mathbf{Y}_{m_Q}$. Each antenna estimates its cross-modal phase, which is used to formulate the loss function observed in Eq.~\eqref{eq:lossfunc}. This loss function is minimized to estimate the misalignment angles $\theta$ and $\phi$, which are then used to compute a phase correction mask. Finally, the phase correction mask is applied on the Rx to restore modal orthogonality and benefit from the multiplexing gain in misaligned cases.

\begin{figure}[t]
    \centering
    \includegraphics[width=0.47\textwidth]{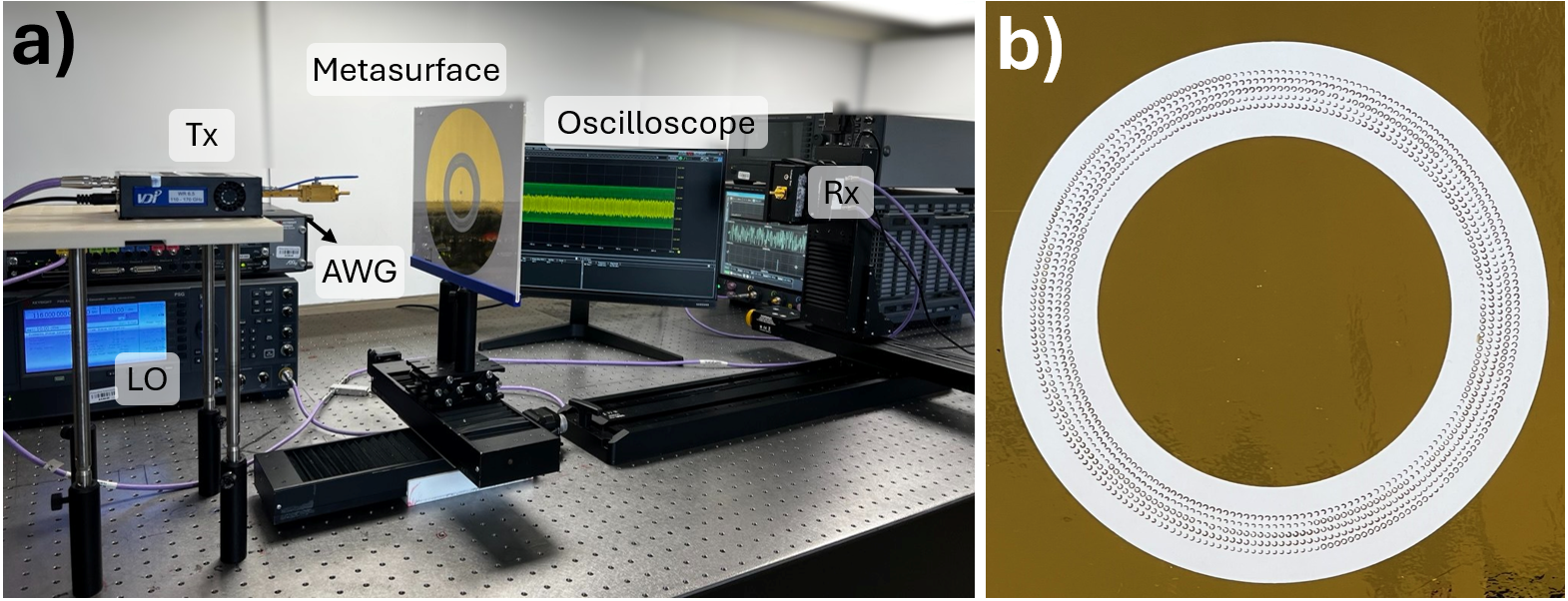}
    \caption{(a) The key components of the setup for over-the-air transmission and reception of OAM beams. (b) The metasurface is designed for the generation of OAM mode $l=1$.}
    \label{fig:setup}
\end{figure}

Before finalizing the optimization framework, we must address a critical question: \textit{How do we select the appropriate sub-carriers, Rx antennas, and Tx modes to use?} Intuitively, the selection process should prioritize sub-carriers, Rx antennas, and Tx modes that provide the highest SNR, to have more robustness against noise.

\textit{Selection of Subcarriers.} In the absence of multipath, all sub-carriers experience nearly identical attenuation according to the free-space path loss equation. This is due to the relatively small bandwidth of the signal compared to its center frequency. Consequently, unlike in multipath-rich environments typical of legacy sub-6 GHz bands, where different sub-carriers can exhibit significantly varying SNRs due to fading, we expect all sub-carriers to have approximately the same SNR in this scenario. As a result, the specific selection of sub-carriers is irrelevant; only the number of sub-carriers used matters. We evaluate the impact of the number of subcarriers on angle estimation accuracy in Section~\ref{sec:NumSC}.

\textit{Selection of Rx Antennas.} In the case of perfect alignment, all antennas in the UCA experience the same SNR due to the radial symmetry of the OAM beam intensity pattern. However, angular tilt can introduce variations in the SNR across the Rx antennas. Unfortunately, these variations are not known in advance at the receiver, as they depend on the misalignment angles, which are yet to be estimated at this point. Since we cannot predict the SNR of each Rx antenna, we do not base our antenna selection on SNR. Instead, we focus on the fact that according to Eq.~\eqref{eq:deltam}, selecting two closely located antennas (with similar $\phi_m$ values) is likely to yield similar cross-modal phase measurements, i.e., less diverse readings. To mitigate this, we adopt a heuristic approach that prioritizes maximizing the spatial separation between the selected antennas. For example, if we have 11 antennas and need to select 3 for angle estimation, we would choose antennas 0, 4, and 8. We evaluate the impact of the number of Rx antennas on angle estimation accuracy in Section~\ref{sec:NumAnt}.

\textit{Selection of Tx Modes.} It is important to note that due to the different divergences of different OAM modes (as discussed in Section~\ref{sec:primer1}), when using a single-ring UCA at both the Tx and Rx, there will always be two modes (symmetric around 0, for example, $l=\pm 2$) that exhibit the highest power compared to the others. The specific modes with this characteristic depend on the approximate distance between the Tx and Rx, as well as their respective aperture sizes, both of which are known in advance. Consequently, we can determine which modes are expected to have the highest SNR at the Rx side, at least in the aligned case. Therefore, it is logical to prioritize these two modes to enhance the SNR for the estimation process. For further discussion on extending this framework to a multi-ring system, refer to Section~\ref{sec:discussion}.

%% file: Sections/Platform.tex
\label{sec:setup}

\noindent
\textbf{Transmission and Reception.} 
The main components of our experimental setup are illustrated in Fig.~\ref{fig:setup}a. We use a Keysight Arbitrary Waveform Generator (M8195A) to generate our intermediate frequency (IF) signal at 4 GHz. A Keysight 306 PSG Analog Signal Generator (E8257D) is then used as the local oscillator (LO) for the up-conversion and down-conversion. The LO is fed into a $4\times$ up-converter from VDI (WR6.5CCU-M4), resulting in a 116 GHz signal. After mixing, the 120 GHz beam is transmitted from the Tx horn antenna to illuminate the transmissive metasurface, at a distance of 14 cm, designed for OAM beam generation. 

The resulting OAM beam propagates in space and is sampled using a single probing antenna. Since no commercial off-the-shelf (COTS) UCA is available at 120 GHz, we emulate one by physically moving the Rx probe to each element’s position. To create 3D misalignment, we rotate the entire virtual array around the $Y$-axis and then around the $X$-axis, thereby assigning new 3D coordinates to each element. The translational stage then moves the probe to these coordinates, sampling the signal as though it were a real, tilted UCA. This setup enables us to experimentally mimic a misaligned array in the absence of the actual hardware.

{We note that although our experimental setup emulates a physical UCA by moving a single probing antenna in 3D space, this does not affect the generalizability of the results. The probing antenna's radiation pattern—though potentially different from a physical UCA element—is captured in the complex gain term $\alpha$ in Eq.~\eqref{Eq:E1}, which models direction-dependent amplitude and phase. For a fixed Rx orientation, all UCA elements receive the signal from the same angle of arrival, so the antenna gain is uniform across the array and acts as a complex scalar. Since \name~uses only the phase differences between OAM modes (i.e., cross-modal phase), this scalar cancels out during processing. Thus, our use of a probing antenna does not impact the accuracy of phase-based angle estimation, and the results remain representative of a real UCA—even under large misalignment—provided the signal lies within the element's field of view.}

For our demonstration, we use a virtual Rx UCA with $N_r = 20$ elements at a radius of $a_r = 8$ mm. Given the physical limitations of the setup and the output power of the transmitter, our experiments are limited to the meter scale. However, the physics of OAM beams and our alignment protocol can be extended to much larger distances. Indeed, past demonstrations have already showcased the potential of OAM beams in the ideal aligned case to achieve a significant data rate of 117 Gbps at a distance of 200 meters~\cite{yagi2021field}.

\begin{figure}[t]
    \centering
    \includegraphics[width=0.47\textwidth]{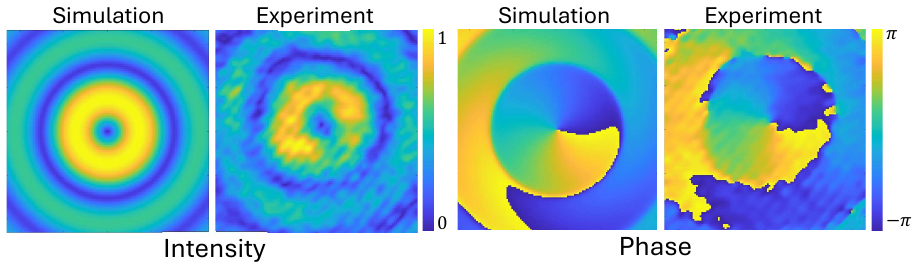}
    \caption{Comparison of the intensity and phase profiles of OAM mode $l=1$ between simulation and measurement results.} 
    \vspace{-0.2cm}
    \label{fig:Heatmap}
\end{figure}

\noindent
\textbf{OAM Generation.} 
OAM beams can be generated using phased arrays~\cite{gong2017generation,liu2016generation}, spiral phase plates~\cite{wei2015generation}, and metasurfaces~\cite{qin2018transmission}. In our setup, we design passive transmissive metasurfaces for OAM beam generation with C-shaped split ring resonators~\cite{shao2024design} as the meta-element. Specifically, we realize the helical phase front in OAM beams by tuning the geometrical features of the meta-element, such as the gap, the opening angle, the width, and the radius. Modifying these parameters allows us to design elements with desired phase shifts such that their phase matches the $e^{il\phi}$ pattern to generate OAM mode $l$. The dimensions of the entire metasurface are 8 by 8 inches. Since each meta-element is extremely small (radius of $0.5$ mm), we create a UCA-like pattern by designing five closely spaced concentric rings, with radii 3.7, 3.85, 4.0, 4.15, and 4.3 cm and element counts 187, 194, 202, 209, and 217, respectively. 

We fabricate our metasurface using a low-cost fabrication method known as hot stamping~\cite{guerboukha2021high}. Specifically, our C-shaped ring resonator patterns are first printed on paper and then laminated with aluminum foil, enabling low-cost, rapid prototyping of metasurfaces. {During experiments, different metasurfaces are substituted to generate different OAM modes.} An example metasurface designed for generating OAM mode 1 is shown in Fig.~\ref{fig:setup}b. 

To verify the metasurface's ability to generate OAM beams, we measured the complex electric field by making the Rx scan a plane perpendicular to the propagation direction at a distance of 40 cm away from the metasurface. We then compare the measured intensity and phase profile to the expected simulated patterns. As shown in Fig.~\ref{fig:Heatmap}, the measured results closely match the simulations, confirming the metasurface's effectiveness in generating OAM beams, despite using a low-cost fabrication technique.

It is important to highlight that our theoretical analysis assumes the far-field condition when modeling the received signal. However, positioning the Rx probe 40~cm from the Tx metasurface places it within the near-field region, since $2D^2/\lambda = 5.12$~m. Remarkably, we observe that \name~remains applicable for this region as well. This robustness arises because the deviation from the far-field assumption introduces a consistent error across all transmitted OAM modes. This uniform error cancels out during the computation of the cross-modal phase, which forms the basis of \name. Consequently, the angle estimation framework operates reliably without modification, even under near-field conditions. We emphasize that this distance was selected based on constraints imposed by the limited link budget and the experimental setup.


{Finally, we emphasize that \name~represents a pioneering effort in experimental OAM within the RF domain. As the hardware technology is not yet mature, COTS UCAs at the D-band, where our experiments are conducted, are unavailable. Consequently, we rely on passive metasurfaces for OAM beam generation. Looking ahead, we envision OAM beams being realized through phased arrays rather than low-cost metasurfaces. Phased arrays will enable dynamic reconfigurability, allowing for rapid switching between OAM modes.}

%% file: Sections/Evaluation.tex
\label{sec:eval}
\subsection{Angle Estimation Performance}
\label{sec:eval_angle}

In this section, we evaluate how well \name~can estimate the misalignment angles by observing the received signal across different OAM modes. To this end, we placed the center of the virtual Rx UCA at a distance of $40$ cm and varied the Rx orientation such that elevation angles ranged from $\theta = 0^\circ$ to $75.7^\circ$ and azimuth angles ranging from $\phi =-180^\circ$ to $-90^\circ$. Please refer to Fig.~\ref{fig:overview} for a visual demonstration of azimuth and elevation angles. 
 
For each pair of angles, the Tx transmits two OAM modes, $l = -1$ and $l = 1$, in two consecutive time slots, meaning the number of modes used in the angle estimation process is $U = 2$. This is achieved by changing the passive Tx metasurface designed for one mode to the other. The phase of the received signal is then measured at all $N_r$ Rx antennas, and a subset of those Rx antennas with size $Q = 6$ is used in the angle estimation process. Note that, for example, if the receiver has three RF chains, necessary to multiplex three independent data streams, \name~becomes a four-shot estimation framework. Specifically, the Tx requires two time slots to transmit the necessary OAM modes, and the Rx requires two time slots to sample the received signals. This reduced time complexity is a key advantage of \name~compared to exhaustive beam-scanning approaches. The cross-modal phase is then estimated and mapped to the estimated elevation and azimuth angles by numerically solving Eq.~\eqref{eq:lossfunc}.  

\begin{figure}[t]
    \centering
    \includegraphics[width=0.47\textwidth]{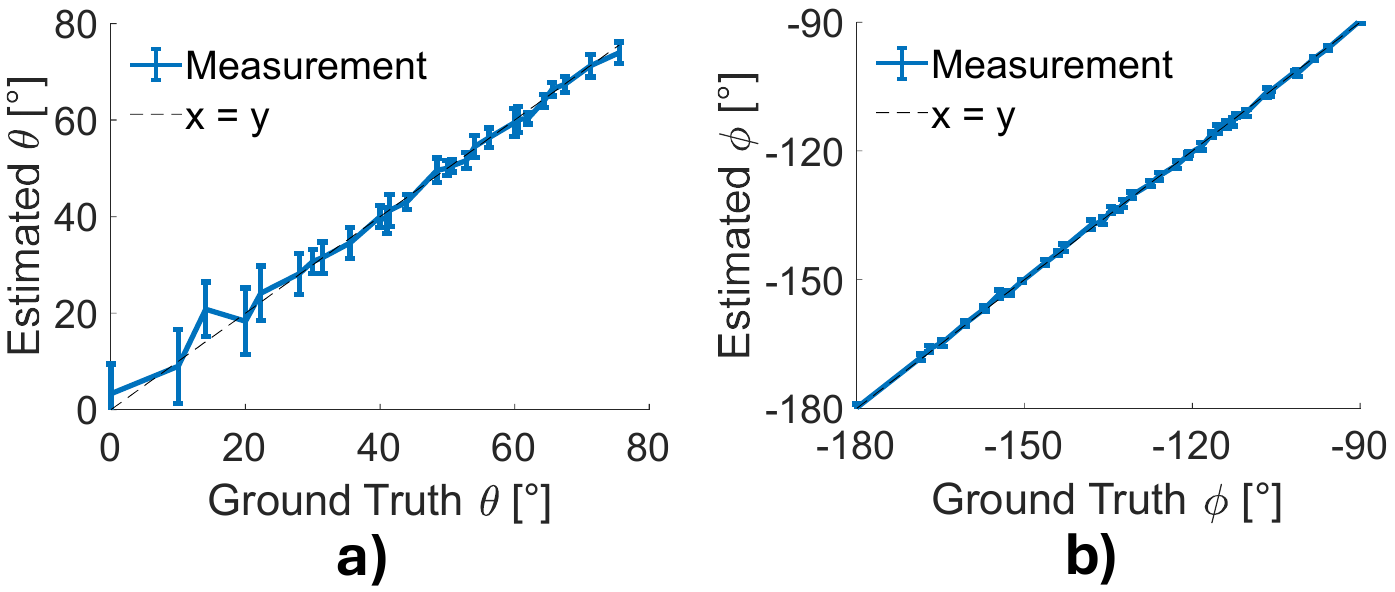}
    \vspace{-0.2cm}
    \caption{\name's estimation of (a) the elevation angle and (b) the azimuth angle using two OAM modes, six antennas, and a single tone at 120 GHz. The error bars denote standard deviation.}
    \label{fig:angle_estimation}
\end{figure}

The results of this experiment are depicted in Fig.~\ref{fig:angle_estimation}, where the average estimated values and their standard deviations (shown as error bars) are plotted for elevation and azimuth angles in Fig.~\ref{fig:angle_estimation}a and Fig.~\ref{fig:angle_estimation}b, respectively. The overall MAE for the estimation of elevation ($\theta$) and azimuth ($\phi$) angles is \ang{2.54} and \ang{0.69}, respectively, demonstrating the accuracy and robustness of \name~in estimating misalignment angles.

These results reveal two interesting observations. First, smaller elevation angles exhibit higher estimation errors, as evident by larger standard deviations in their estimates in Fig.~\ref{fig:angle_estimation}a. To explain this, recall that our angle estimation framework relies on estimating $\delta_m(\theta,\phi)$ for different antennas and using it to solve for $\theta$ and $\phi$. When $\delta_m$ changes more rapidly with respect to $\theta$ or $\phi$, it serves as a more effective signature for angle inference. As seen in Eq.~\eqref{eq:deltam}, when the elevation angle is small, $\delta_m(\theta,\phi)$ is less sensitive to changes in $\theta$ because the derivative of $\cos\theta$ is small. Consequently, small variations in $\theta$ produce minimal changes in $\delta_m$, making it harder to resolve small elevation angles accurately. In contrast, larger elevation angles induce more pronounced changes in $\delta_m$, improving estimation accuracy.  

Second, we observe that the overall estimation accuracy is higher for the azimuth angle compared to the elevation angle, as indicated by its smaller MAE. This can also be attributed to the relative sensitivity of $\delta_m$ with respect to $\phi$ and $\theta$ within the range of angles tested in our experiments.

\textit{Overall, \name~can precisely identify the misalignment angles with MAE of \ang{2.54} and \ang{0.69} for the elevation and azimuth angles, respectively, by just using two OAM modes and six receiver antennas at a single frequency.}

\subsection{Restoring Modal Orthogonality}
\label{eval:SIR}

\begin{figure}[t]
    \centering
    \includegraphics[width=0.47\textwidth]{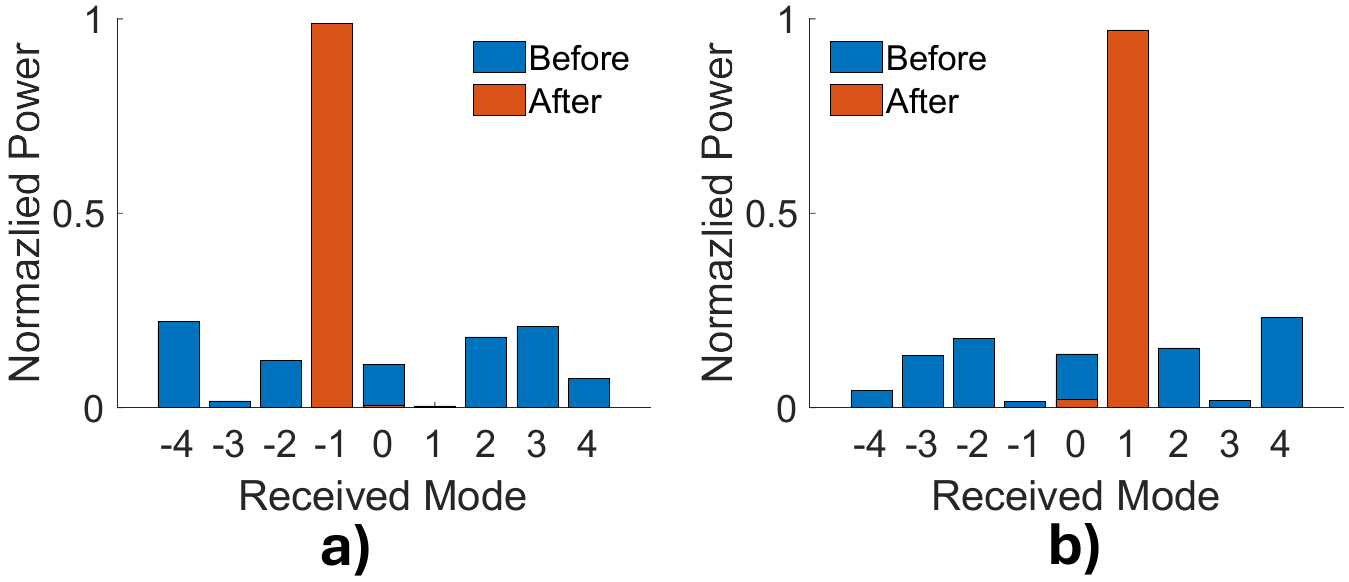}
    \vspace{-0.2cm}
    \caption{The IMI patterns before and after applying \name~are shown for when the Tx is transmitting mode (a) $l=-1$ and (b) $l=1$.} 
    \label{fig:IMI}
\end{figure}

In this section, we evaluate the effectiveness of \name~in mitigating IMI. Hence, here, we calculate and apply the phase correction masks at each Rx misalignment setting according to the estimated angle results reported in Section~\ref{sec:eval_angle}. In other words, the error in our angle estimation would yield an imperfect phase correction mask, which itself results in residual IMI. 

To provide a visual illustration of how IMI is resolved by using \name's framework, we plotted an example map of leaked energy from the transmitted mode $l=-1$ and $1$ before and after the proposed phase correction. In this setting, the Rx was misaligned by elevation angle $\theta = 14.3^\circ$ and azimuth angle $\phi = -134.7^\circ$. The results are depicted in Fig.~\ref{fig:IMI}. We observe that before \name~applies the phase correction mask, the energy from the transmitted mode leaks to other modes, causing IMI (similar to simulation results in Fig.~\ref{fig:misalignment}b). Yet, after the phase correction mask is applied, the power in the correct mode significantly increases, while other modes are suppressed (similar to Fig.~\ref{fig:misalignment}c). 

\begin{figure}[t]
    \centering
    \includegraphics[width=0.47\textwidth]{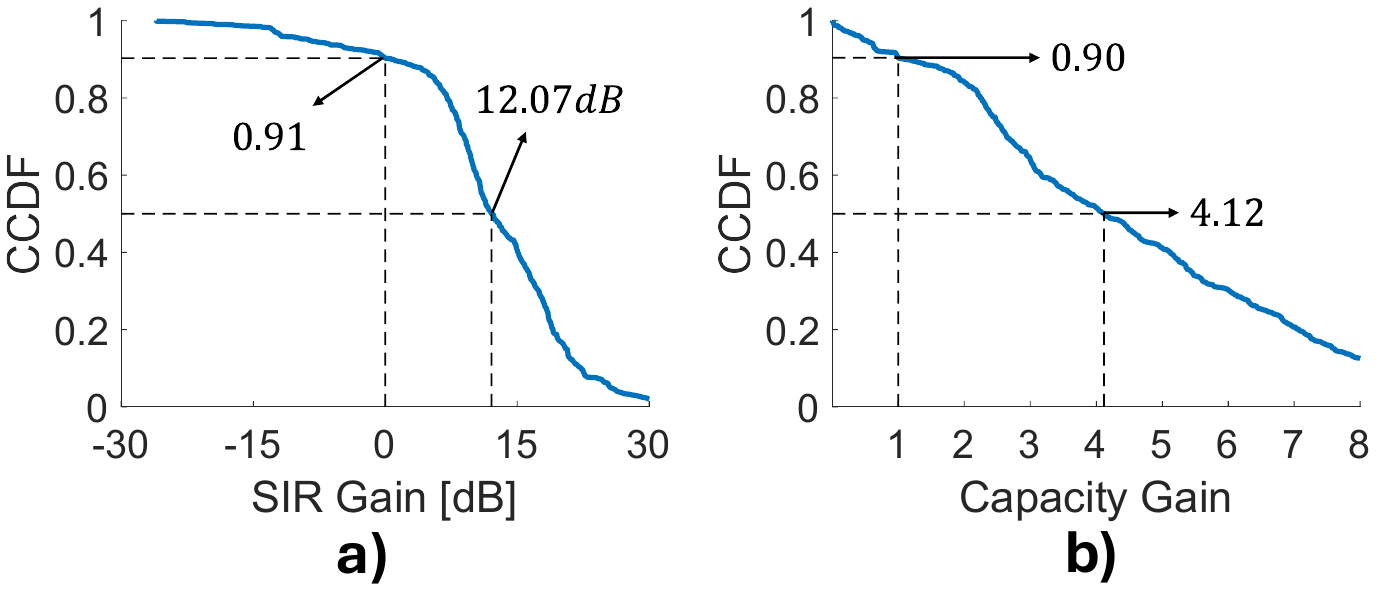}
    \vspace{-0.2cm}
    \caption{CCDF of (a) the SIR gain and (b) the capacity gain by using \name.}
    \label{fig:SIR}
\end{figure}

Using the same experimental setup as Section~\ref{sec:eval_angle}, we computed the average SIR before and after applying the phase mask across the two transmitted modes (-1 and 1). We define the difference between these values as SIR gain, which quantifies the improvement achieved in reducing IMI by using \name. The results, depicted in Fig.~\ref{fig:SIR}a as an empirical complementary cumulative distribution function (CCDF), show that \name~achieves an average SIR gain of 12.08 dB.  

To provide more insights into how much gain \name~provides in terms of channel capacity, we also calculated the Shannon channel capacity, before and after applying the phase mask calculated by \name. We assumed we are in the low-noise regime, meaning the dominant term in reducing the channel capacity is interference. The results are plotted in the form of empirical CCDF in Fig.~\ref{fig:SIR}b. We observe an average $4.57\times$ gain in the channel capacity when using \name~to restore mode orthogonality.

\textit{\name~can effectively reduce the inter-modal interference, boosting the SIR by more than 12 dB, and improving the channel capacity by more than 4.5-fold, all while demanding only a few-shot measurement.}

\subsection{Impact of Number of Subcarriers}
\label{sec:NumSC}

\begin{figure}[t]
    \centering
    \includegraphics[width=0.47\textwidth]{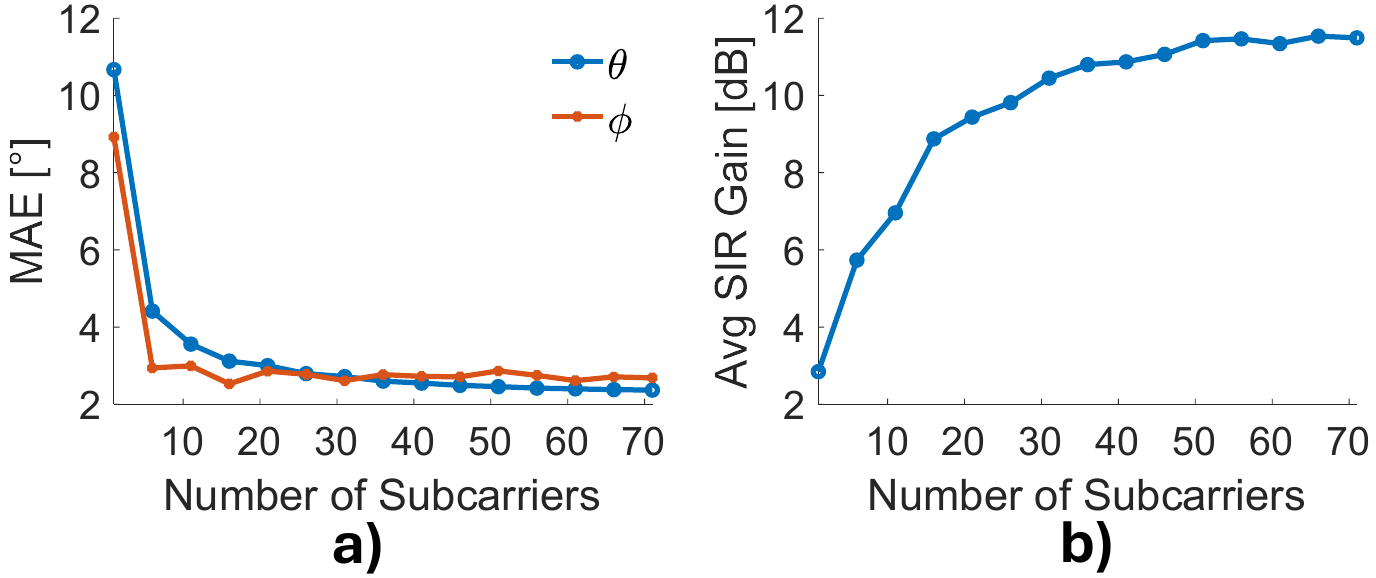}
    \vspace{-0.2cm}
    \caption{Increasing the number of subcarriers improves (a) the accuracy of misalignment angle estimation and (b) the average SIR gain.}
    \label{fig:NumSC}
\end{figure}

So far, we have seen the performance of \name~in inferring the Rx orientation via assessing the cross-modal phase seen at one frequency tone. Next, we evaluate how increasing the total bandwidth can enhance the estimation framework in \name~and improve the SIR. As discussed in Section~\ref{sec:framework}, using more subcarriers can improve angle estimation performance because the cross-modal phase is invariant to frequency and can be measured across any frequency. By increasing the bandwidth and measuring the cross-modal phase at more frequencies, we can average out noise and improve estimation accuracy.  

To test this, we used the same geometrical configuration as in Section~\ref{sec:eval_angle}, but this time transmitted a multi-tone signal spanning 119.5 to 120.2 GHz with a spacing of 10 MHz, resulting in a total of 71 subcarriers. We note that the frequency response of the C-shaped ring resonators can be assumed to be flat in this bandwidth. We then used a random subset of these subcarriers of size $P$, to estimate the misalignment angles $\theta$ and $\phi$. By varying $P$, we analyzed the impact of the number of subcarriers on estimation accuracy. In Fig.~\ref{fig:NumSC}a, we have plotted the MAE of the estimated angles as a function of the number of sub-carriers included in the process. This result shows that increasing the number of subcarriers reduces the MAE of the estimated angles, confirming that leveraging frequency diversity is effective.  

Additionally, we calculated the average improvement in SIR after applying the phase mask computed from the estimated angles, as a function of the number of subcarriers used. These results, plotted in Fig.~\ref{fig:NumSC}b, show that leveraging a wider range of frequencies leads to greater SIR improvement, further validating the benefits of frequency diversity in \name. Furthermore, this additional accuracy comes at no extra time overhead, but at the cost of bandwidth, which is assumed to be readily available in these high-frequency regimes. 

\textit{\name~leverages the fact that the cross-modal phase in OAM reception is inherently frequency-independent. By utilizing information from multiple subcarriers within a bandwidth of less than 1 GHz, \name~improves Rx angle estimation and, consequently, the SIR.}

\subsection{Impact of  Number of Antennas}
\label{sec:NumAnt}

\begin{figure}[t]
    \centering
    \includegraphics[width=0.47\textwidth]{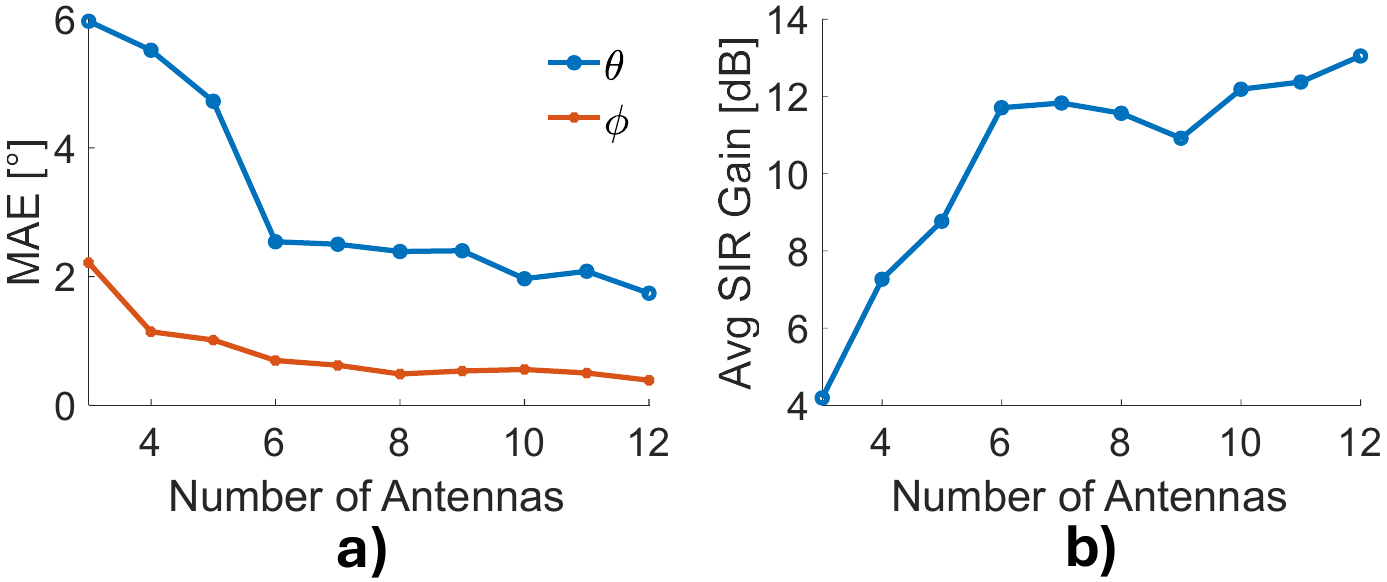}
    \vspace{-0.2cm}
    \caption{Using more Rx antennas enhances (a) the precision of misalignment angle estimation and (b) the average gain in SIR.} 
    \vspace{-0.2cm}
    \label{fig:NumAnt}
\end{figure}
Recall that in Section \ref{sec:framework}, we discussed using more Rx antennas as another method to improve the accuracy of our angle estimation framework. The key idea is that by increasing the number of antennas used in the estimation process, we reduce the dependency of the estimated angles on any single measurement of the cross-modal phase. This redundancy helps mitigate the impact of noise and measurement errors, leading to more robust angle estimation.  

In this section, we evaluate this claim using the same parameters as in Section~\ref{sec:eval_angle}, but varying the number of antennas included in the estimation process from 3 to 12. Note that 3 is the minimum number required to estimate the three unknowns: $\theta$, $\phi$, and $\gamma$. We plot the MAE of the angle estimation in Fig.~\ref{fig:NumAnt}a and the average SIR gain in Fig.~\ref{fig:NumAnt}b as functions of the number of antennas used. We observe that increasing the number of antennas is indeed an effective method for reducing angle estimation errors, leading to higher improvements in the SIR. Notably, as shown in the plot, most of the improvement is achieved using 6 antennas, with performance gains plateauing beyond this point. This is a critical insight because, unlike increasing the number of subcarriers (as discussed in Section~\ref{sec:NumSC}), increasing the number of antennas would linearly increase the time overhead of \name. Therefore, this parameter must be chosen carefully to balance accuracy and time efficiency. Fortunately, based on our results, using six antennas seems to be the sweet spot in balancing the trade-off between accuracy and time complexity of our angle estimation method. 

\textit{\name~can improve the robustness to noise by utilizing the redundancy in Rx antennas; however, only a few antennas are enough to yield accurate estimations. }

%% file: Sections/Discussions.tex
\label{sec:discussion}

\noindent 
\textbf{OAM vs. Beamforming.}
In all MIMO systems, a fundamental question arises: should one employ beamforming or spatial multiplexing? Given that OAM mode multiplexing is a specialized instance of spatial multiplexing (see Section~\ref{sec:primer2}), the well-established comparisons between spatial multiplexing and beamforming naturally extend to the case of OAM mode multiplexing versus beamforming. Theoretically, if transmit power is allocated to each OAM mode using the water-filling principle, OAM mode multiplexing always achieves a higher overall communication rate than beamforming. However, this advantage depends on system geometry, specifically, the distance between arrays relative to their aperture size and the noise level. In low SNR regimes, where only one OAM mode has significant gain, water-filling naturally concentrates power in that mode, making the solution effectively equivalent to beamforming. In such cases, the performance gap between OAM mode multiplexing and beamforming is negligible, while beamforming remains simpler to implement. Thus, although OAM mode multiplexing is theoretically superior, beamforming is preferred in low SNRs due to its practicality.

\noindent
\textbf{Additional Link-Level Limitations.}
{In this paper, we addressed Tx–Rx misalignment as a key barrier to the practical deployment of OAM links. However, other challenges remain, including limited aperture due to beam divergence and susceptibility to multipath. These issues fall outside the scope of this work and present opportunities for future research.}

\noindent 
\textbf{Translational Misalignment.}
In this work, we examine a scenario in which the Tx actively tracks the receiver Rx, and any misalignment is attributed solely to angular offsets at the Rx. Under these conditions, only two angles, elevation and azimuth, need to be estimated to compute the phase correction mask. However, in a more general context where both Tx and Rx may be misaligned independently, the estimation must extend to four angles: elevation and azimuth at both the Rx and the Tx. Such a scenario demands a cooperative estimation framework, where both ends iteratively adjust their phase masks to restore orthogonality. A key challenge emerges when severe misalignment causes the Tx beam to miss the Rx aperture entirely, leading to insufficient SNR for angle estimation, an issue that presents an interesting direction for future work.

\noindent 
\textbf{Long-Range Operation.}
We acknowledge that the experiments were conducted over a short range (40 cm). However, this limitation was due to low Tx power and constraints in the setup. Prior work has shown OAM links at much higher distances~\cite{yagi2021field}. Crucially, \name~can be extended to such ranges without modifcations to the framework.

\noindent 
\textbf{Dynamic-Arrays.}
In this work, we used passive metasurfaces to generate OAM beams, since configurable UCAs are not commercially available in this band. However, these arrays have been fabricated in industry labs~\cite{hirabe201940,sasaki2023demonstration}. The next step for \name~is to implement it on dynamic arrays. Yet, we emphasize that this transition does not require any changes to the protocol.

\noindent 
\textbf{Alternate Array Geometries.}
Our current framework models both the Tx and Rx as single-ring UCAs. However, other array geometries, such as uniform concentric circular arrays (UCCAs) and uniform rectangular arrays, can also generate OAM beams. Among these, UCCAs are particularly promising because, in a multi-ring system, we can design the radii of rings on the Tx side for transmitting specific OAM modes, as well as on the Rx side for receiving them. By designing the radii and modes appropriately, a multi-ring system can ensure sufficient power for all modes, enabling the use of mode diversity to improve angle estimation performance, as discussed in Section~\ref{sec:framework}. Thus, a follow-up for \name~is to adapt it to these geometries.

%% file: Sections/Conclusion.tex
\label{sec:conclusion}
We presented \name, a novel framework addressing the fundamental challenge of misalignment that disrupts modal orthogonality and diminishes multiplexing gains in OAM communications. \name~achieves accurate inference of the misalignment angles via a few-shot measurement of the received signal under various transmitted OAM modes, providing precise angle estimation with an MAE of \ang{2.54} for the elevation angle and \ang{0.69} for the azimuth angle. This angle estimation is used to restore the orthogonality by applying a phase correction mask at the Rx, improving the SIR by more than 12 dB and increasing the channel capacity by a factor of 4.5, all without requiring power-hungry and costly fully digital arrays. Our experimental validation using low-cost metasurfaces at 120 GHz demonstrates \name’s robustness in real-world conditions, bridging the gap between theoretical OAM advantages and practical use cases. {Overall, \name~paves the way for enhancing the throughput of LOS OAM wireless networks through real-time misalignment estimation and compensation, enabling more robust operation in applications such as wireless backhauling and satellite-to-ground communication.}

%% file: OAM.bbl
\begin{thebibliography}{10}
\providecommand{\url}[1]{#1}
\csname url@samestyle\endcsname
\providecommand{\newblock}{\relax}
\providecommand{\bibinfo}[2]{#2}
\providecommand{\BIBentrySTDinterwordspacing}{\spaceskip=0pt\relax}
\providecommand{\BIBentryALTinterwordstretchfactor}{4}
\providecommand{\BIBentryALTinterwordspacing}{\spaceskip=\fontdimen2\font plus
\BIBentryALTinterwordstretchfactor\fontdimen3\font minus \fontdimen4\font\relax}
\providecommand{\BIBforeignlanguage}[2]{{%
\expandafter\ifx\csname l@#1\endcsname\relax
\typeout{** WARNING: IEEEtran.bst: No hyphenation pattern has been}%
\typeout{** loaded for the language `#1'. Using the pattern for}%
\typeout{** the default language instead.}%
\else
\language=\csname l@#1\endcsname
\fi
#2}}
\providecommand{\BIBdecl}{\relax}
\BIBdecl

\bibitem{telatar1999capacity}
E.~Telatar, ``{Capacity of multi-antenna Gaussian channels},'' \emph{European transactions on telecommunications}, vol.~10, no.~6, pp. 585--595, 1999.

\bibitem{mahmouli20134}
F.~E. Mahmouli and S.~D. Walker, ``{4-Gbps uncompressed video transmission over a 60-GHz orbital angular momentum wireless channel},'' \emph{IEEE Wireless Communications Letters}, vol.~2, no.~2, pp. 223--226, 2013.

\bibitem{ren2017line}
Y.~Ren, L.~Li, G.~Xie, Y.~Yan, Y.~Cao, H.~Huang, N.~Ahmed, Z.~Zhao, P.~Liao, C.~Zhang \emph{et~al.}, ``{Line-of-sight millimeter-wave communications using orbital angular momentum multiplexing combined with conventional spatial multiplexing},'' \emph{IEEE Transactions on Wireless Communications}, vol.~16, no.~5, pp. 3151--3161, 2017.

\bibitem{willner2021orbital}
A.~E. Willner, K.~Pang, H.~Song, K.~Zou, and H.~Zhou, ``{Orbital angular momentum of light for communications},'' \emph{Applied Physics Reviews}, vol.~8, no.~4, 2021.

\bibitem{trichili2019communicating}
A.~Trichili, K.-H. Park, M.~Zghal, B.~S. Ooi, and M.-S. Alouini, ``{Communicating using spatial mode multiplexing: Potentials, challenges, and perspectives},'' \emph{IEEE Communications Surveys \& Tutorials}, vol.~21, no.~4, pp. 3175--3203, 2019.

\bibitem{yagi2021field}
Y.~Yagi, H.~Sasaki, T.~Semoto, T.~Kageyama, T.~Yamada, J.~Mashino, and D.~Lee, ``{{Field experiment of 117 Gbit/s wireless transmission using OAM multiplexing at a distance of 200 m on 40 GHz band}},'' in \emph{{2021 IEEE International Conference on Communications Workshops (ICC Workshops)}}.\hskip 1em plus 0.5em minus 0.4em\relax IEEE, 2021, pp. 1--5.

\bibitem{lee2018experimental}
D.~Lee, H.~Sasaki, H.~Fukumoto, Y.~Yagi, T.~Kaho, H.~Shiba, and T.~Shimizu, ``{An experimental demonstration of 28 GHz band wireless OAM-MIMO (orbital angular momentum multi-input and multi-output) multiplexing},'' in \emph{{2018 IEEE 87th Vehicular Technology Conference (VTC Spring)}}.\hskip 1em plus 0.5em minus 0.4em\relax IEEE, 2018, pp. 1--5.

\bibitem{sasaki2023demonstration}
H.~Sasaki, Y.~Yagi, R.~Kudo, and D.~Lee, ``{Demonstration of 1.44 Tbit/s OAM multiplexing transmission in sub-THz bands},'' in \emph{{2023 IEEE International Conference on Communications Workshops (ICC Workshops)}}.\hskip 1em plus 0.5em minus 0.4em\relax IEEE, 2023, pp. 338--343.

\bibitem{hirabe201940}
M.~Hirabe, R.~Zenkyu, H.~Miyamoto, K.~Ikuta, and E.~Sasaki, ``{40 m transmission of OAM mode and polarization multiplexing in E-band},'' in \emph{{2019 IEEE Globecom Workshops (GC Wkshps)}}.\hskip 1em plus 0.5em minus 0.4em\relax IEEE, 2019, pp. 1--6.

\bibitem{yang2018digital}
B.~Yang, Z.~Yu, J.~Lan, R.~Zhang, J.~Zhou, and W.~Hong, ``{Digital beamforming-based massive MIMO transceiver for 5G millimeter-wave communications},'' \emph{IEEE Transactions on Microwave Theory and Techniques}, vol.~66, no.~7, pp. 3403--3418, 2018.

\bibitem{zhu2014demystifying}
Y.~Zhu, Z.~Zhang, Z.~Marzi, C.~Nelson, U.~Madhow, B.~Y. Zhao, and H.~Zheng, ``{Demystifying 60GHz outdoor picocells},'' in \emph{{Proceedings of the 20th annual international conference on Mobile computing and networking}}, 2014, pp. 5--16.

\bibitem{li2018experimental}
L.~Li, R.~Zhang, G.~Xie, Y.~Ren, Z.~Zhao, Z.~Wang, C.~Liu, H.~Song, K.~Pang, R.~Bock \emph{et~al.}, ``{Experimental demonstration of beaconless beam displacement tracking for an orbital angular momentum multiplexed free-space optical link},'' \emph{Optics Letters}, vol.~43, no.~10, pp. 2392--2395, 2018.

\bibitem{xie2015exploiting}
G.~Xie, L.~Li, Y.~Ren, Y.~Yan, N.~Ahmed, Z.~Zhao, Z.~Wang, N.~Ashrafi, S.~Ashrafi, R.~D. Linquist \emph{et~al.}, ``{Exploiting the unique intensity gradient of an orbital-angular-momentum beam for accurate receiver alignment monitoring in a free-space communication link},'' in \emph{{2015 European Conference on Optical Communication (ECOC)}}.\hskip 1em plus 0.5em minus 0.4em\relax IEEE, 2015, pp. 1--3.

\bibitem{xie2017localization}
G.~Xie, L.~Li, Y.~Ren, Y.~Yan, N.~Ahmed, Z.~Zhao, C.~Bao, Z.~Wang, C.~Liu, H.~Song \emph{et~al.}, ``{Localization from the unique intensity gradient of an orbital-angular-momentum beam},'' \emph{Optics Letters}, vol.~42, no.~3, pp. 395--398, 2017.

\bibitem{pang2020experimental}
K.~Pang, H.~Song, X.~Su, K.~Zou, Z.~Zhao, H.~Song, A.~Almaiman, R.~Zhang, C.~Liu, N.~Hu \emph{et~al.}, ``{Experimental mitigation of the effects of the limited size aperture or misalignment by singular-value-decomposition-based beam orthogonalization in a free-space optical link using Laguerre--Gaussian modes},'' \emph{Optics Letters}, vol.~45, no.~22, pp. 6310--6313, 2020.

\bibitem{cui2024effect}
X.~Cui, K.-H. Park, and M.-S. Alouini, ``{Effect of Random Misalignment in the Capacity of Millimeter-wave OAM},'' \emph{IEEE Open Journal of the Communications Society}, vol.~5, pp. 1141--1154, 2024.

\bibitem{chen2018beam}
R.~Chen, H.~Xu, M.~Moretti, and J.~Li, ``{Beam steering for the misalignment in UCA-based OAM communication systems},'' \emph{IEEE Wireless Communications Letters}, vol.~7, no.~4, pp. 582--585, 2018.

\bibitem{zhengjuan2021broadband}
T.~Zhengjuan, C.~Rui, L.~Wenxuan, Z.~Hong, and M.~Marco, ``{Broadband beam steering for misaligned multi-mode OAM communication systems},'' \emph{Journal of Systems Engineering and Electronics}, vol.~32, no.~4, pp. 779--788, 2021.

\bibitem{chen2019reception}
R.~Chen, W.-X. Long, and J.~Li, ``{Reception of misaligned multi-mode OAM signals},'' in \emph{{2019 IEEE Global Communications Conference (GLOBECOM)}}.\hskip 1em plus 0.5em minus 0.4em\relax IEEE, 2019, pp. 1--5.

\bibitem{yu2022uca}
W.~Yu, B.~Zhou, Z.~Bu, and Y.~Zhao, ``{UCA based OAM beam steering with high mode isolation},'' \emph{IEEE Wireless Communications Letters}, vol.~11, no.~5, pp. 977--981, 2022.

\bibitem{saito2021efficient}
S.~Saito, Y.~Ito, H.~Suganuma, K.~Ogawa, and F.~Maehara, ``{Efficient inter-mode interference cancellation method for OAM multiplexing in the presence of beam axis misalignment},'' in \emph{{2021 IEEE International Conference on Communications Workshops (ICC Workshops)}}.\hskip 1em plus 0.5em minus 0.4em\relax IEEE, 2021, pp. 1--6.

\bibitem{cheng2019achieving}
W.~Cheng, H.~Jing, W.~Zhang, Z.~Li, and H.~Zhang, ``{Achieving practical OAM based wireless communications with misaligned transceiver},'' in \emph{{ICC 2019-2019 IEEE International Conference on Communications (ICC)}}.\hskip 1em plus 0.5em minus 0.4em\relax IEEE, 2019, pp. 1--6.

\bibitem{jian2021non}
M.~Jian, Y.~Chen, and G.~Yu, ``{Non-coaxial OAM: Precoding design, misaligned parameter estimation and capacity compensation},'' \emph{IEEE Access}, vol.~9, pp. 37\,726--37\,738, 2021.

\bibitem{chen2023index}
M.~Chen, R.~Chen, Y.~Zhao, Z.~Yang, and Y.~L. Guan, ``{Index-modulation OAM detectors resistant to beam misalignment},'' \emph{IEEE Transactions on Vehicular Technology}, vol.~73, no.~2, pp. 2836--2841, 2023.

\bibitem{vahidiniai2021array}
V.~Vahidiniai, M.~Atashbar, and S.~Hosseinzadeh, ``{Array misalignment angle value estimation for OAM-based communication system},'' \emph{Journal of the Franklin Institute}, vol. 358, no.~18, pp. 10\,213--10\,231, 2021.

\bibitem{tian2016beam}
H.~Tian, Z.~Liu, W.~Xi, G.~Nie, L.~Liu, and H.~Jiang, ``{Beam axis detection and alignment for uniform circular array-based orbital angular momentum wireless communication},'' \emph{Iet Communications}, vol.~10, no.~1, pp. 44--49, 2016.

\bibitem{sun2023enhanced}
J.-J. Sun, S.~Sun, L.-J. Yang, and J.~Hu, ``{Enhanced misalignment estimation of orbital angular momentum signal based on deep recurrent neural networks},'' \emph{IEEE Transactions on Antennas and Propagation}, vol.~71, no.~6, pp. 5522--5527, 2023.

\bibitem{chen2020multi}
R.~Chen, W.-X. Long, X.~Wang, and L.~Jiandong, ``{Multi-mode OAM radio waves: Generation, angle of arrival estimation and reception with UCAs},'' \emph{IEEE Transactions on Wireless Communications}, vol.~19, no.~10, pp. 6932--6947, 2020.

\bibitem{long2021aoa}
W.-X. Long, R.~Chen, M.~Moretti, and J.~Li, ``{AoA estimation for OAM communication systems with mode-frequency multi-time ESPRIT method},'' \emph{IEEE Transactions on Vehicular Technology}, vol.~70, no.~5, pp. 5094--5098, 2021.

\bibitem{gao2019misalignment}
X.~Gao, X.~Song, Z.~Zheng, M.~Xie, and S.~Huang, ``{Misalignment measurement of orbital angular momentum signal based on spectrum analysis and image processing},'' \emph{IEEE Transactions on Antennas and Propagation}, vol.~68, no.~1, pp. 521--526, 2019.

\bibitem{allen1992orbital}
L.~Allen, M.~W. Beijersbergen, R.~Spreeuw, and J.~Woerdman, ``{Orbital angular momentum of light and the transformation of Laguerre-Gaussian laser modes},'' \emph{Physical review A}, vol.~45, no.~11, p. 8185, 1992.

\bibitem{mohammadi2009orbital}
S.~M. Mohammadi, L.~K. Daldorff, J.~E. Bergman, R.~L. Karlsson, B.~Thid{\'e}, K.~Forozesh, T.~D. Carozzi, and B.~Isham, ``{Orbital angular momentum in radio—A system study},'' \emph{IEEE transactions on Antennas and Propagation}, vol.~58, no.~2, pp. 565--572, 2009.

\bibitem{gong2017generation}
Y.~Gong, R.~Wang, Y.~Deng, B.~Zhang, N.~Wang, N.~Li, and P.~Wang, ``{Generation and transmission of OAM-carrying vortex beams using circular antenna array},'' \emph{IEEE Transactions on antennas and propagation}, vol.~65, no.~6, pp. 2940--2949, 2017.

\bibitem{bhardwaj2019generation}
S.~Bhardwaj, ``{Generation and manipulation of oam beams in 2d planar arrays and reflectarrays},'' in \emph{{2019 International Applied Computational Electromagnetics Society Symposium (ACES)}}.\hskip 1em plus 0.5em minus 0.4em\relax IEEE, 2019, pp. 1--2.

\bibitem{wei2015generation}
X.~Wei, C.~Liu, L.~Niu, Z.~Zhang, K.~Wang, Z.~Yang, and J.~Liu, ``{Generation of arbitrary order Bessel beams via 3D printed axicons at the terahertz frequency range},'' \emph{Applied optics}, vol.~54, no.~36, pp. 10\,641--10\,649, 2015.

\bibitem{qin2018transmission}
F.~Qin, L.~Wan, L.~Li, H.~Zhang, G.~Wei, and S.~Gao, ``{A transmission metasurface for generating OAM beams},'' \emph{IEEE Antennas and Wireless Propagation Letters}, vol.~17, no.~10, pp. 1793--1796, 2018.

\bibitem{edfors2011orbital}
O.~Edfors and A.~J. Johansson, ``{Is orbital angular momentum (OAM) based radio communication an unexploited area?}'' \emph{IEEE Transactions on Antennas and Propagation}, vol.~60, no.~2, pp. 1126--1131, 2011.

\bibitem{gil2021comparison}
G.-T. Gil, J.~Y. Lee, H.~Kim, and D.-H. Cho, ``{Comparison of UCA-OAM and UCA-MIMO systems for sub-THz band line-of-sight spatial multiplexing transmission},'' \emph{Journal of Communications and Networks}, vol.~23, no.~2, pp. 83--90, 2021.

\bibitem{starin2011attitude}
S.~R. Starin and J.~Eterno, ``{Attitude determination and control systems},'' \emph{NASA Technical Reports}, 2011.

\bibitem{liu2016generation}
K.~Liu, H.~Liu, Y.~Qin, Y.~Cheng, S.~Wang, X.~Li, and H.~Wang, ``{Generation of OAM beams using phased array in the microwave band},'' \emph{IEEE Transactions on Antennas and Propagation}, vol.~64, no.~9, pp. 3850--3857, 2016.

\bibitem{shao2024design}
Z.~Shao, H.~Chen, R.~Shen, Y.~Ghasempour, and K.~Sengupta, ``{Design of Metasurfaces for Beamforming with High-Efficiency Custom Polarization Generation},'' in \emph{{2024 IEEE International Symposium on Antennas and Propagation and INC/USNC-URSI Radio Science Meeting (AP-S/INC-USNC-URSI)}}.\hskip 1em plus 0.5em minus 0.4em\relax IEEE, 2024, pp. 1159--1160.

\bibitem{guerboukha2021high}
H.~Guerboukha, Y.~Amarasinghe, R.~Shrestha, A.~Pizzuto, and D.~M. Mittleman, ``{High-volume rapid prototyping technique for terahertz metallic metasurfaces},'' \emph{Optics Express}, vol.~29, no.~9, pp. 13\,806--13\,814, 2021.

\end{thebibliography}
